\documentclass[12pt]{article}

\usepackage{epsfig}

\newcommand{\lp}{\left(}
\newcommand{\rp}{\right)}

\usepackage{amsmath,amssymb}

\def\simlt{\stackrel{<}{{}_\sim}}
\def\simgt{\stackrel{>}{{}_\sim}}

\newcommand{\be}{\begin{equation}}
\newcommand{\ee}{\end{equation}}
\newcommand{\bea}{\begin{eqnarray}}
\newcommand{\eea}{\end{eqnarray}}

\newcommand{\dsVs}{\partial_{s_r} {\tilde V}_1|_s}

\newcommand{\dVbs}{\Delta {\tilde V}_{1bs}}
\newcommand{\Vob}{{\tilde V}_{1}|_b}

\newcommand{\deh}{\partial_{h_r}}
\newcommand{\dhh}{\partial^2_{h_r} }
\newcommand{\dhs}{\partial^2_{h_rs_r}}
\newcommand{\ds}{\partial_{s_r} }
\newcommand{\dss}{\partial^2_{s_r} }
\newcommand{\dsss}{\partial^3_{s_r} }

\numberwithin{equation}{section}

\begin{document}

\baselineskip=18pt

\setcounter{footnote}{0}
\setcounter{figure}{0}
\setcounter{table}{0}

\begin{titlepage}
\begin{flushright}
CERN-PH-TH/2011-171\\
\today \\
\end{flushright}
\vspace{.3in}

\begin{center}
{\Large \bf Strong Electroweak Phase Transitions\\ 
in the Standard Model with a Singlet}

\vspace{0.5cm}

{\bf J.R. Espinosa$^{a,b}$, T. Konstandin$^c$} and {\bf F. Riva$^b$}

\centerline{$^{a}${\it ICREA, Instituci\'o Catalana de Recerca i Estudis 
Avan\c{c}ats,}}
\centerline{$^b${\it  IFAE, Universitat Aut\`onoma de Barcelona, 08193 
Bellaterra, Barcelona, Spain}}
\centerline{$^{c}${\it Physics Department, CERN, CH--1211 Geneva 23, 
Switzerland}}

\end{center}
\vspace{.8cm}

\begin{abstract}
\medskip
\noindent
It is well known that the electroweak phase transition (EWPhT) in extensions of the Standard Model with one real scalar singlet can be first-order for realistic values of the Higgs mass. We revisit this scenario with the most general renormalizable scalar potential systematically identifying
 all regions in parameter space that develop, due to tree-level dynamics, a potential barrier at the critical temperature that is strong enough to avoid sphaleron wash-out of the baryon asymmetry. Such strong EWPhTs allow for a simple mean-field approximation and an analytic treatment of the free-energy that leads to very good theoretical control and understanding of the different mechanisms that can make the transition strong. We identify a new realization of such   mechanism, based on a flat direction developing at the critical temperature, which could operate in other models. Finally, we discuss in detail some special cases of the model performing a  numerical calculation of the one-loop free-energy that improves over the mean-field approximation and confirms the analytical expectations. 
\end{abstract}

\bigskip
\bigskip

\end{titlepage}


\section{Introduction}

The search for physics beyond the Standard Model (SM) has strong
theoretical and experimental motivation and the simplest 
extension is to enhance the SM by a scalar gauge singlet degree of
freedom. This minimalistic model (and its cousins with a complex
singlet or supersymmetric versions of it) can be very successful in explaining various
phenomena that cannot be explained by the SM: dark
matter~\cite{Bento:2000ah}-\cite{Ponton}, spontaneous $B-L$
breaking~\cite{Majoron}-\cite{EWPHTMajoron} and the baryon asymmetry of
the Universe~\cite{Huber:2000mg, EWPHTnMSSM}, often leading to characteristic collider signatures~\cite{Datta:1995qw}-\cite{Gripaios:2009pe}.

One prominent difference between the SM and its singlet extensions is the following. While in the SM  the LEP bound on the Higgs mass ($M_h>114.4$ GeV \cite{LEPMh}) implies that the electroweak phase transition (EWPhT) is not first-order but a smooth crossover \cite{crossover}, the addition of a singlet can lead to strongly first-order  EWPhTs~\cite{AH}-\cite{Ashoorioon:2009nf}
for realistic values of the scalar masses. Moreover, with such strong EWPhTs, not only the
observed baryon asymmetry of the Universe can be explained through electroweak (EW) baryogenesis  (provided the model also contains additional sources of $CP$ violation) but the process
of EW symmetry breaking can also leave the trace of a  stochastic background of gravitational waves~\cite{Apreda:2001us}.

The aim of the present work is to revisit the EWPhT in the most general
renormalizable extension of the SM with one additional real scalar
singlet. Although this issue has been studied in the past~\cite{AH}-\cite{Ashoorioon:2009nf},
(both numerically and analytically at different levels of generality), we believe that a thorough  \emph{analytical} understanding of the rich spectrum of possibilities for a strong EWPhT this model offers is still lacking in the literature. The analysis that comes closest to this task is Ref.~\cite{Ramsey}, over which we will improve in a number of aspects. 

In the SM and some extensions of it, a first-order EWPhT 
is caused by the thermal effects of bosons coupled to the Higgs,
that generate a cubic term in the Higgs scalar potential. Although this can be successful in many cases, it requires sizeable couplings of these bosons to the Higgs and the effect can be screened by thermal masses when
daisy resummation is taken into account. In this article, we concentrate on EWPhTs for which the barrier separating broken and symmetric vacua is not generated by the previous thermal cubic correction but rather by tree-level effects. These tree-level effects lead in general to stronger EWPhTs as $v_c=v(T_c)$, 
the Higgs vacuum expectation value (VEV) at the critical temperature $T_c$ (that controls through the celebrated ratio $v_c/T_c$ the sphaleron erasure of the baryon asymmetry), is now proportional to some $T$-independent dimensionful parameter in the potential; hence $v_c/T_c$ can become potentially very large for small $T_c$.\footnote{Strong EWPhTs are particularly welcome if, as suggested by \cite{DeSimone}, magnetic fields generated at the EWPhT lower the sphaleron energy so that larger values of $v_c/T_c$ are required to avoid baryon washout.}
The parameter space of this model is quite rich and we will see that these tree-level barriers are not necessarily related to the presence of cubic terms in the potential, as is often assumed.

We begin in Section~\ref{sec:PhT} by studying the structure of the tree-level scalar potential of the model. In particular we are interested in potentials where the EW breaking and preserving minima are degenerate, since this is the situation that arises for strong phase transitions. Indeed, in this case it is well justified to use the mean-field approximation for the free-energy, which will have the same structure as the tree-level potential (with temperature-dependent parameters).
Differently from previous analyses, we will introduce a novel set of parameters particularly convenient for the discussion of the vacuum structure of this kind of potentials (and which might also be of use for phenomenological studies of the scalar sector of these models). In spite of the large number of free parameters (eight) we have to deal with, this new parametrization will allow us to identify very easily, and analytically, the structure of the potential: its stable minima and the existence of a barrier between them.  As a result, beside developing a better understanding of the ingredients necessary to get
a strong EWPhT in this model, we will find new scenarios with strong EWPhTs that had not been identified before (involving in particular flat directions at the critical temperature).

In Section~\ref{sec:strong_PT} we discuss thermal corrections to the scalar potential and explain our strategy to search for regions in parameter space with strong EWPhTs, which we summarize in Table~\ref{Table1}. The idea is to start from a potential with degenerate broken and unbroken minima and a barrier between them: this is identified with a potential at $T_c$ which gives a strong EWPhT. Its parameters are then evolved to lower $T$ to find their values at $T=0$, where they can be used for phenomenological purposes. After identifying the regions in parameter space that give a strong EWPhT, we then perform a more precise analysis  including one-loop corrections (at $T=0$ and finite temperature)  without resorting to high-$T$ expansions and including daisy resummation. Although the strength of the EWPhT is somewhat reduced with respect to tree-level, one still gets sizable values for it. Our results confirm then the expectations based on the tree-level analysis.

In the rest of the paper we apply these tools to particular realizations of the model  previously considered in the literature: the $\mathbf{Z}_2$-symmetric
case in Section~\ref{sec:Z2}; a particular supersymmetric incarnation in Section~\ref{sec:nmssm}; and a case with a very light scalar in Section~\ref{sec:dark}. Finally, we study some examples of the general case in Section~\ref{sec:general} and conclude in Section~\ref{sec:conclusions}. Appendix~\ref{app:thermalpot} contains some technical details of the full one-loop analysis, including the $T=0$ renormalization conditions.

\section{Tree-level Scalar Potential\label{sec:PhT}}

\subsection{Parametrization of the Potential}

To begin with, we look for a convenient parametrization of the
potential that ensures control over its minima: this will allow us to easily identify which are the global minima and whether or not they are stable. The most general (renormalizable) tree-level
potential for the SM Higgs field $h$ and the singlet $s$
depends on 8 parameters,
\be
\label{Vtree}
V= -\frac{1}{2}\mu_h^2 h^2+\frac{1}{4}\lambda_h h^4 +\frac{1}{2}\mu_s^2 s^2 +\frac{1}{4}\lambda_s s^4+
\frac{1}{4}\mu_m s h^2+\frac{1}{4}\lambda_m s^2h^2 +\mu_1^3 s+\frac{1}{3}\mu_3 s^3\ .
\ee
Note that a
redefinition of the singlet field by a constant shift, $s\rightarrow s
+ \sigma$, simply amounts to a redefinition of the  parameters
$\mu_1^3, \mu_s^2, \mu_h^2, \mu_3$ and $\mu_m$ but does not change the physics,
being just a coordinate change:
\bea
\mu_1^3 &\rightarrow &\mu_1^3 + \lambda_s \sigma^3+\mu_3\sigma^2+\mu_s^2\sigma\nonumber\ , \\
\mu_s^2 &\rightarrow &\mu_s^2 + 3\lambda_s \sigma^2+2\mu_3\sigma\nonumber\ , \\
\mu_h^2 &\rightarrow &\mu_h^2 - \frac{1}{2}\lambda_m \sigma^2-\frac{1}{2}\mu_m\sigma\nonumber\ , \\
\mu_3 &\rightarrow &\mu_3 + 3\lambda_s \sigma\nonumber\ , \\
\mu_m &\rightarrow &\mu_m + 2\lambda_m \sigma\ .
\label{shifts}
\eea
This shift is often used to get rid of
one of the initial parameters, choosing e.g. $\mu_1=0$ or $\mu_3=0$.\footnote{If the potential is invariant under the discrete $\mathbf{Z}_2$ symmetry $s\rightarrow -s$ then $\mu_m=\mu_1=\mu_3=0$ is the best "coordinate frame" for the singlet as it makes explicitly apparent such symmetry.}
However, different
choices can be useful in different circumstances so we refrain from any particular choice at this stage and 
keep the discussion as general as possible. Still, it is of advantage to
choose a parametrization in which the shift symmetry is realized in a
more explicit way.  Beside this property, the new parameters will allow for a more direct theoretical control of
the structure of the potential. The parameters we introduce are: the
vacuum expectation values $v\equiv\langle h\rangle$ and 
$w\equiv\langle s\rangle$ in the broken minimum; the three elements 
of the scalar squared-mass matrix,
evaluated at the broken minimum (as indicated by the subscript $b$),
\be
m_{h}^2\equiv\left.\frac{\partial^2 V}{\partial h\partial h}\right|_b\ ,\;\;
m_{s}^2\equiv\left.\frac{\partial^2 V}{\partial s\partial s}\right|_b\ ,\;\;
m_{sh}^2\equiv\left.\frac{\partial^2 V}{\partial h\partial s}\right|_b\ ;\;\;
\label{masses}
\ee
the mixed quartic coupling $\lambda_m$; the effective coupling
$\lambda^2$ defined by
\be
\lambda^2 \equiv \lambda_h\lambda_s-\frac{1}{4}\lambda_m^2\ ,
\ee
that appears recurrently in different contexts (note that $\lambda^2$ can be negative); and finally, the combination 
\be
m_* = \lambda^2w + \frac{1}{3}\lambda_h\mu_3-\frac{1}{8}\lambda_m \mu_m\ ,
\ee
which can be checked to be shift-invariant.

To sum up, our parameters are 
\be
\label{eq:params1}
\{ v, w, m_{h}^2, m_{s}^2, m_{sh}^2, \lambda_m, \lambda^2, m_*\}\ .
\ee
With the exception of $w$ (transforming as $w\rightarrow w-\sigma$), all these parameters are shift-independent.
For reasons that will become clear when we discuss the thermal
potential, it is convenient not to assume at this point that $v$ takes
its standard value $v_{EW}=246$~GeV. To avoid confusion later on, we
reserve the notation $v_{EW}$ for the latter value. The relations
between the old and the new parameters are:
\bea
\label{reli}
\mu_h^2 & = & \frac{1}{2}m_h^2+\frac{w}{v}m_{sh}^2-\frac{1}{2}\lambda_m w^2\ , \\
\mu_s^2 & = & m_s^2-\frac{1}{2}\lambda_m v^2
+ \frac{3 v w}{2 m_h^2} \left[ -2 \lambda_m m_{sh}^2 - 8 m_* v 
+ (4 \lambda^2 + \lambda_m^2) v w \right]\ , \\
\mu_3 & = & \frac{3v}{2m_h^2}\left[4m_*v+\lambda_m m_{sh}^2
-(4\lambda^2+\lambda_m^2)v w\right]\ ,\\
\mu_m&=& 2\frac{m_{sh}^2}{v}-2\lambda_m w\ ,\label{mum}\\
\mu_1^3 & = & \frac{v}{2}(\lambda_m v w -m_{sh}^2)-m_s^2w 
+ \frac{v w^2}{2 m_h^2} \left[ 3 \lambda_m m_{sh}^2 
+ 12 m_* v - (4 \lambda^2 + \lambda_m^2) v w\right]\ , \\
\lambda_h & = & \frac{1}{2}\frac{m_h^2}{v^2}\ ,\\
\lambda_s & = &  (4\lambda^2+\lambda_m^2)\frac{v^2}{2m_h^2}\ .
\label{relf}
\eea
This change of variables is non-singular: indeed its Jacobian is simply $12 (m_s^2m_h^2-m_{sh}^4)/m_h^4$ which, as shown in the next Section [eq.~(\ref{eq:detM})], is always positive and non-singular.

The potential in the new parametrization reads (up to an appropriate constant)
\bea
V &=& \frac{ m_h^2 }{8 v^2} \lp h^2 - v^2 \rp^2  
+  \frac{m_{sh}^2}{2 v} \lp h^2 - v^2 \rp (s - w) 
+ \frac14 \left[ 2 m_s^2 +\lambda_m(h^2  - v^2) \right] (s - w)^2 \nonumber\\ 
& + & \frac1{2 m_h^2} (\lambda_m m_{sh}^2 + 4 m_* v) v (s - w)^3  
  + \frac{1}{8 m_h^2} (4 \lambda^2 + \lambda_m^2) v^2 (s - w)^4\ ,
\label{newpot}
\eea
where we have expressed it as a polynomial in $(s-w)$, showing explicitly that now the parameter $w$ can be used to absorb any shift in $s$, leaving the other parameters of the potential invariant.

\subsection{Structure of the Potential}

As a good starting point for the analysis, we would like our
tree-level potential to be well-behaved. First, it should have a stable broken-phase minimum $(v,w)$. This is guaranteed by using $v$ and $w$ as input
parameters and by a judicious choice of the mass-matrix elements in
(\ref{masses}), such that both mass eigenvalues are real and positive,
that is,
\be
\label{eq:detM}
{\mathrm{Det}}\ {\cal{M}}^2_s=
m_{s_1}^2m_{s_2}^2=m_s^2m_h^2-m_{sh}^4>0\ .
\ee
Alternatively, one can use directly the mass eigenvalues,
$m_{s_1}^2, m_{s_2}^2$ and the scalar mixing angle, $\alpha_{sh}$, as input
parameters, imposing the relevant experimental bounds \cite{Hbounds} (suitably modified to take the singlet mixing into account).

\subsubsection{\em Stability\label{stability}}

We also want that the tree-level potential does not have
unbounded-from-below directions. The large-field behaviour of the potential
is dominated by the quartic part, which in our parametrization reads:
\be\label{V444}
V_4=\frac{1}{8m_h^2v^2}\left[(m_h^2 h^2+\lambda_m v^2 s^2)^2+4\lambda^2v^4 s^4
\right]\ .
\ee
For $\lambda_m<0$, the squared-sum term vanishes along the directions $h=\pm\sqrt{-\lambda_m }v s /m_h$, so that $\lambda^2>0$ is required to ensure stability. For $\lambda_m>0$, on the other hand, the squared-sum term is always positive and the positivity condition on $\lambda^2$ can be relaxed but, to maintain stability along the $s$-direction, one should require $\lambda^2>-\lambda_m^2/4$ in that case [so that $\lambda_s>0$ in the parametrization of eq.~(\ref{Vtree})].
 
The fact that $(v,w)$ is a local minimum does not guarantee that it is the global one: a deeper minimum might exist. We will derive below the
necessary and sufficient conditions for this situation to arise. These conditions turn out to be extremely simple in terms of
our new parameters.

\subsubsection{\em Local Minima at $\mathbf{h\neq 0}$.}

\begin{figure}[t]
\includegraphics[width=0.95\textwidth,clip]{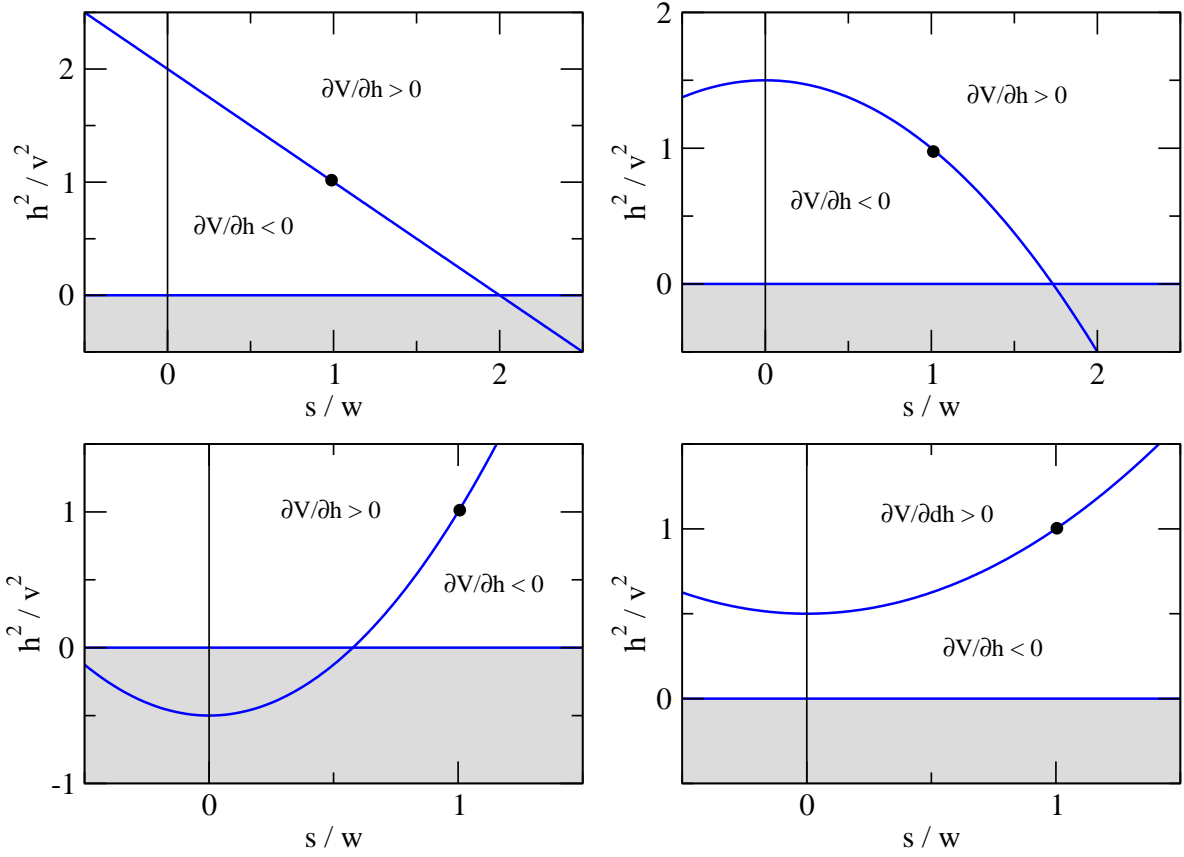}
\caption{
\label{fig:types} 
The curve $h^2=D^2_h(s)$ in the $(h^2/v^2,s/w)$-plane. The different cases
correspond to: $\lambda_m=0$ (upper left); $\lambda_m>0$ (upper right); $\lambda_m<0$
with $|\lambda_m|<m_{sh}^4/(m_h^2v^2)$ (lower left);  $\lambda_m<0$ with
$|\lambda_m|>m_{sh}^4/(m_h^2v^2)$ (lower right). The unphysical region $h^2<0$ is shaded gray. The EW breaking minimum at $h^2=v^2$ and $s=w$ is marked by a black dot.}
\end{figure}

Now we search for further local minima of the potential in order to
ensure that $(v,w)$ is the global one or, at least, degenerate with an unbroken minimum (as relevant  for the EWPhT). 
The stationary points of the
tree-level potential lie on the  curves along which 
$\partial V/\partial h=0$,
\bea
\label{h0h1}
\frac{\partial V}{\partial h}=0 &\Rightarrow &\left\{h=0\ ,\;\;\;\;
 {\rm and } \;\;\;\;
h^2=D^2_h(s)\equiv \frac{1}{2\lambda_h}(2\mu_h^2-\mu_m s-\lambda_m s^2)\right\}\ .\ 
\eea
In our parametrization, the curve $D_h^2(s)$ reads
\be
\label{h12s}
D^2_h(s)=v^2  -2 v (s-w)\frac{m_{sh}^2}{m_h^2}-\frac{\lambda_m v^2}{m_h^2}(s-w)^2\ .
\ee
In particular, we have $D^2_h(w)=v^2$ (by construction) and it is
interesting that this curve is independent of the 
parameters $m_*$, $m_s^2$ and $\lambda^2$. Its shape, determined by $\lambda_m$ and $m_{sh}^2$, will be relevant later on. Fig.~\ref{fig:types} shows the different possibilities for the $D^2_h(s)$ and $h=0$ lines in the $(h^2/v^2,s/w)$-plane, with a minimum at $s=w$ as indicated. These lines separate the
plane in regions of definite sign of $\partial V/\partial h$. 

For $\lambda_m=0$, the curve $D_h^2(s)$ is a straight line, intersecting the axis $h=0$ at one single point (Fig.~\ref{fig:types}, upper left).
For the special case in which also $m_{sh}^2=0$ one simply
has $D^2_h(s)=v^2$, and the corresponding line is parallel to $h=0$.

In general, with $\lambda_m\neq 0$, $D_h^2(s)$ is a
parabola and can have three qualitatively different forms. If
$\lambda_m>0$, $D_h^2(s)$ curves down and intersects $h=0$ at two
points. If $\lambda_m<0$, $D_h^2(s)$ curves up. For
$|\lambda_m|>m_{sh}^4/(m_h^2 v^2)$, $D_h^2(s)$  does not intersect $h=0$
while, in the opposite case, it has two intersection points. For either
sign of $\lambda_m$, the two intersection points with the $s$-axis are given by
\be
\label{spm}
s_{\pm}-w = \frac{1}{\lambda_m  v}\left[-m_{sh}^2  \pm 
\sqrt{m_{sh}^4+\lambda_m v^2 m_h^2}\right]\ .
\ee
All these different possibilities are illustrated in Fig.~\ref{fig:types}. We can use the shift freedom to move the axis of the parabola to $s=0$, which corresponds to enforcing  
\be
w = \frac{m_{sh}^2}{\lambda_m  v},
\ee
or to setting $\mu_m=0$ in the original parametrization.
This is the choice we generally adopt in our plots.

Next, we consider possible additional stationary points along the curve
$h^2=D^2_h(s)$. The potential along such curve, $V[D_h(s),s]$, is a
quartic potential in $s$ that can be minimized in a straightforward
manner.  More explicitly, the minimization equation $d V[D_h(s),s]/d
s=0$, leads in the general case to a cubic equation of the form
\be
a (s-w)^3 + b (s-w)^2 + c (s-w) + d = 0\ ,
\ee
with 
\be
a  =  2 \lambda^2 v^2, \,\,\,\,
b  =  6 m_* v^2, \,\,\,\,
c  =  {\rm Det} {\cal M}_s^2, \,\,\,\,
d  =  0\ .
\ee
The nature and number of real solutions this cubic equation has is
determined, as usual, by the discriminant
\be
\Delta=18abcd-4b^3d+b^2c^2-4ac^3-27a^2d^2\ .
\ee
For $\Delta<0$ there is only one real root, corresponding to a single
minimum, the electroweak one; for $\Delta >0$ there are three real
roots (the previous minimum and two other stationary points); 
for $\Delta=0$ the two additional roots
merge in an inflection point.  Notice that these additional stationary
points are only physically relevant if they appear in the region with
$D_h^2(s)>0$ (the interval $[s_-,s_+]$, with $s_{\pm}$ defined by
eq.~(\ref{spm}), if $\lambda_m>0$; or the intervals $[-\infty,s_-]$,
$[s_+,\infty]$ if $\lambda_m<0$). With our coordinates,
\be\label{signDELTA}
\mathrm{sign}(\Delta)=
\mathrm{sign}\left[ 9 m_*^2 v^2 - 2 \lambda^2 {\rm Det} {\cal M}_s^2 \right] \ .
\ee
Recall that ${\mathrm{Det}}\ {\cal{M}}^2_s=m_s^2m_h^2-m_{sh}^4>0$ from
(\ref{eq:detM}). This means that the necessary condition to have
an additional stationary point along $D^2_h(s)$ is 
\be
\label{extrastat}
\lambda^2< \tilde \lambda^2 \equiv \frac{9 m_*^2 v^2}{ 2 {\rm Det} {\cal M}_s^2} \ ,
\ee
where $\tilde\lambda^2>0$.

In the case $0<\lambda^2<\tilde \lambda^2$, the two additional
stationary points are another minimum and a maximum separating it from
the EW breaking one.  Their location is also easy to obtain: they
appear at $(D_h(w_\pm),w_\pm)$ with
\bea\label{wpiumeno}
w_\pm - w \equiv 
-\frac{3 m_*}{2 \lambda^2} 
 \pm \frac{1}{2 \lambda^2 v}
\sqrt{9 m_*^2 v^2 - 2 \lambda^2 {\rm Det} {\cal M}_s^2}\ .
\eea
By evaluating the potential at these points, it is straightforward to
obtain the condition for the minimum at $(v,w)$ and the additional one from eq.~(\ref{wpiumeno}) to be degenerate:\footnote{In this degenerate case,
$w_- - w = 2(w_+ - w)$, corresponding to a potential $V[D_h(s),s]$
symmetric under $(s-w_+)\rightarrow -(s-w_+)$ (this is generic for a
quartic potential with two degenerate minima).}
\be\label{lambdaPot2min}
\lambda^2 = 8 \tilde \lambda^2 /9\ .
\ee

For $\lambda^2<0$ [which requires $\lambda_m>0$ from the stability discussion below eq.~(\ref{V444})], the two additional stationary points are
two maxima, with the EW minimum between them\footnote{In this case $V[D_h(s),s]$ is
unbounded from below for $s\rightarrow\pm\infty$; however, this is not a
problem because this region is not physical when $\lambda_m>0$, as discussed above eq.~(\ref{signDELTA}).}.  The EW
minimum will still be the deepest (physically relevant) minimum along
$D^2_h(s)$ provided $V(v,w)<V(0,s_\pm)$. If this is not the case, a
deeper minimum must exist at $h=0$. We discuss such situation in the next Subsection.

\begin{figure}[t]
\includegraphics[width=0.95\textwidth, clip ]{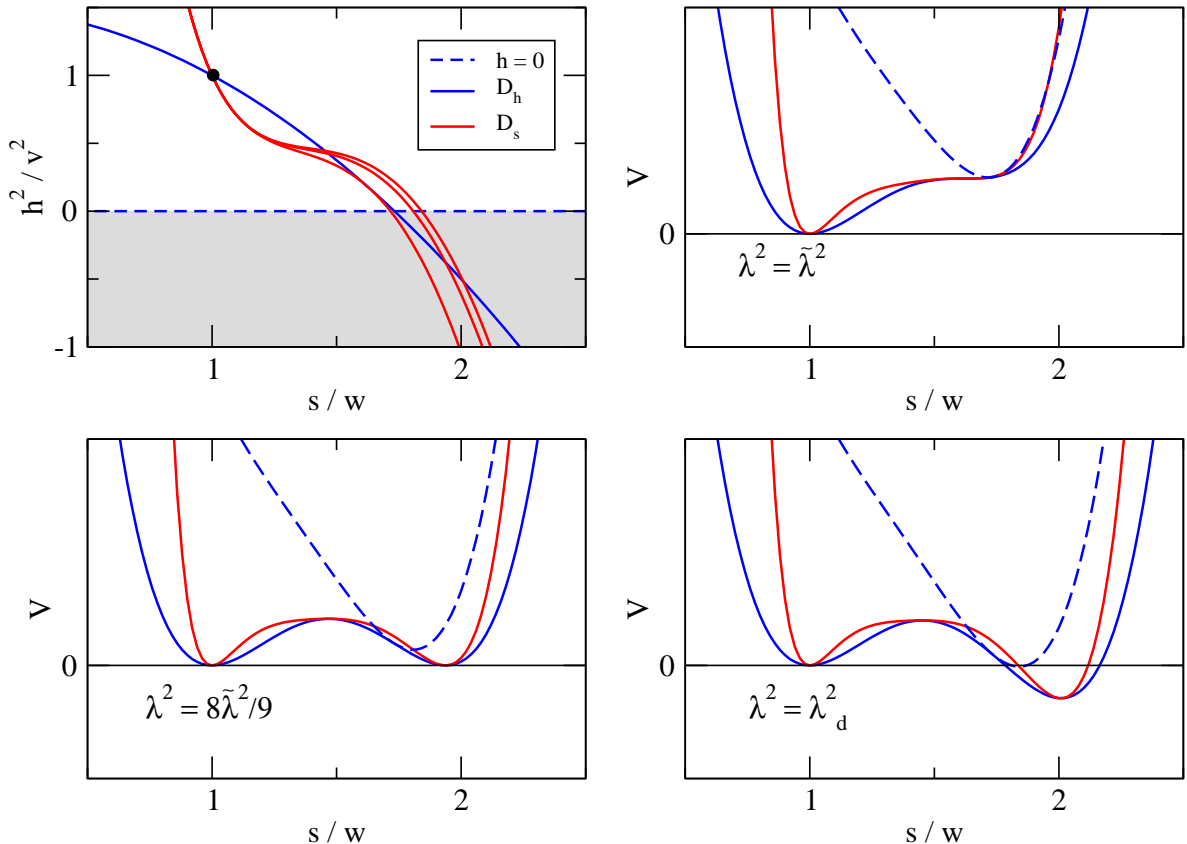}
\caption{\label{fig:mins} Example for the dependence of the potential
on $\lambda^2$. The upper left plot shows $D^2_h(s)$ (solid blue) and $D^2_s(s)$ [along which $\partial V/\partial s=0$, see eq.~(\ref{h22})] (solid red) for several
values of $\lambda^2$: $\tilde\lambda^2$, $8\tilde\lambda^2/9$ and $\lambda_d^2$. The intersections of these two curves correspond to the stationary points of the potential. The remaining plots show the potential along
$D^2_h(s)$, $D^2_s(s)$ (same color coding) and $h=0$ (dashed) at the indicated values of $\lambda^2$.}
\end{figure}

To summarize, as illustrated by Fig.~\ref{fig:mins}, for a potential with all parameters fixed except
$\lambda^2$, the possible stationary points away from $h=0$ lie along
a fixed curve $D^2_h(s)$, independent of $\lambda^2$, with a minimum
located at $(v,w)$ by construction. For large enough $\lambda^2>\tilde\lambda^2>0$, the
minimum at $(v,w)$ is the only stationary point. When $\lambda^2=\tilde \lambda^2$ an inflection point
develops, while for smaller $\lambda^2$ there are two minima. For $\lambda^2=8 \tilde \lambda^2/9$, the new
non-standard minimum is degenerate with the one at $(v,w)$ and for smaller $\lambda^2$ our minimum $(v,w)$ is no longer the lowest one. Whether these other minima are
physically relevant or not will depend on whether they appear at
positive values of $D_h^2(s)$ or not.

\subsubsection{\em Local Minima at $\mathbf{h= 0}$.\label{localminima0}}

To discuss the possible presence of minima in the direction $h=0$,
which might be deeper than the electroweak vacuum, two simple facts
are relevant. First, it is useful to note the following relation:
\be
\label{ineq}
V[D_h(s),s]-V[0,s]= -\frac{m_h^2}{8v^2} D_h^4(s) < 0\ .
\ee
A glance at Fig.~\ref{fig:types} explains this fact as a result of the sign of $\partial V/\partial h$ in the region between the lines $h=0$ and $h^2=D^2_h(s)$. 
Second,
\be
\label{mass0}
\left.\frac{\partial^2 V[h,s]}{(\partial h)^2}\right|_{h=0} =
-\frac{m_h^2}{2v^2}  D_h^2(s)\ .  
\ee
This implies that minima along $h=0$ can only appear in regions with
$D_h^2(s)<0$.\footnote{An alternative way to see this is to note that $\partial^2V/(\partial h)^2=0$ along the line $h^2=D_h^2(s)/3$, which divides the  $(h^2,s)$-plane in two regions with opposite signs of $\partial^2V/(\partial h)^2$.}  Moreover, in order to locate such minima, it is enough
to minimize $V[0,s]$: if a minimum found in this way appears at
$D_h^2(s)<0$, then (\ref{mass0}) ensures that it is also a minimum
along the $h$-direction. We also conclude that, for fixed $s$
the potential can have only one minimum: at $h=0$ for $D_h^2(s)<0$
or at $h\neq0$ for $D_h^2(s)>0$.

More can be said about these minima at $h=0$ by considering the following
general statement about the tree-level potential: \emph{If there are two local minima
with $h^2>0$ then there is no local minimum with $h=0$}. To prove this,
assume there are two local minima with $h^2>0$ and one with $h=0$, chosen 
to be the absolute minimum along $h=0$ and to lie at $s=0$.
Consider a curve of the form
\be
h^2=D^2(s) \equiv \alpha s + \beta s^2=\beta s (s-s_1)\ ,
\ee
that passes through all three minima (two minima with
equal $s$ are not possible and so, this curve always exists). The
potential along this curve, $V_D(s)\equiv V[D(s),s]$, is a quartic polynomial 
and, therefore, can have at most three stationary points.  Two of them are
the two minima with $h^2>0$ while the minimum at $h=0$ is in general not
a stationary point of $V_D(s)$. The reason is that
$dV[D(s),s]/ds = [dV/d(h^2)]  dD^2(s)/ds + dV/ds$, but $dV/dh = 0$ does not in general imply  $dV/d(h^2) = 0$ because $h=0$ leads to $dV/dh=0$ even if $dV/d(h^2)\neq 0$.\footnote{A relevant exception to this is $h^2=D_h^2(s)$ as then we always have $dV/d(h^2)=0$ by construction.} Furthermore, minima must 
be separated by maxima and in most geometrical arrangements of the locations
of the three minima this will require more stationary points than the allowed maximum 
of three. The only non-trivial case occurs if the minimum at $h=0$ is between
the other two (ordered by their $s$-coordinates) and $s_1\neq 0$, but in that case it is straightforward to see that $V_D(s_1)<V_D(0)$, contradicting our assumption
that $s=0$ is the absolute minimum along $h=0$, and this concludes the proof.\footnote{Obviously, this "theorem" applies to the tree-level potential only and can be violated through loop corrections. It is nevertheless  useful in order to identify large barriers created by tree-level effects.}

On the other hand, {\em if
there is only one local minimum with $h^2>0$ there can be up to two
local minima with $h=0$}. The case with two minima at $h=0$ requires that the potential $V_D(s)$
has one minimum (the one corresponding to $h\neq 0$) and is negative for $s \to \pm\infty$. In addition, the two minima at $h=0$ lie at both sides of the EW minimum. (An explicit
example will be given in the $\mathbf{Z}_2$ symmetric case below).

Having these facts in mind, we can compare minima at $h=0$ with our minimum $(v,w)$ and discuss what are the conditions on the parameters of the potential for $(v,w)$ to be
the global minimum. We will illustrate this with an example in Fig.~\ref{fig:mins}, which plots the potential along $h^2=D^2_h(s)$ and $h=0$ for different values of $\lambda^2$. Consider first the case in which 
$(v,w)$ is the deepest minimum of $V[D_h(s),s]$ (i.e. we have
$\lambda^2>8\tilde \lambda^2/9$), then (\ref{ineq}) immediately implies
it should also be the global minimum of $V[h,s]$.  Such case is shown in Fig.~\ref{fig:mins}, upper right plot. 
Cases with $\lambda^2 <8
\tilde \lambda^2/9$, for which a deeper minimum along $D^2_h(s)$
appears (or, if $\lambda^2<0$,  when $V[D_h(s),s]$ is unbounded from below), might still have $(v,w)$ as the global minimum (this happens if the
new minimum is in the unphysical region, $D_h^2(s)<0$), as in
Fig.~\ref{fig:mins}, lower left plot.  In such cases one needs to check the minima
along $h=0$, which might be deeper than the EW one without violating
(\ref{ineq}).    As $\lambda^2$ gets more
and more negative, minima along $h=0$ might become the global minimum.  In general, when all parameters except
$\lambda^2$ are fixed, the potential along $h=0$ decreases with
decreasing $\lambda^2$ [see the explicit potential in eq.~(\ref{newpot})]. 
Hence, there is a definite value $\lambda_d^2$ for
$\lambda^2$ that separates the region in parameter space in which the
EW minimum is the deepest one from the one where it is not. It is
clear that $-\lambda_m^2/4 \leq \lambda_d^2 \leq
8\tilde\lambda^2/9$. The exact value of $\lambda_d^2$ \label{promisedlambda} will 
be determined below by requiring degeneracy between the EW minimum and
the second minimum [see eq.~(\ref{lambda2deg})], an example of which is shown in the lower right plot of Fig.~\ref{fig:mins}. 
Studying such degenerate cases will be very relevant
for the discussion of strong phase transitions in the next Section, 
so we turn to this issue next.

\subsection{Coexisting and Degenerate Minima.}

The most interesting cases for the phase transition study are potentials with two degenerate minima: the EW-breaking one, at $(v,w)$, and  the symmetric one at $(0,w_0)$. While we could use the shift of eq.~(\ref{shifts}) to specify the value of the singlet field VEV in the unbroken phase, $w_0$ ($w_0=0$ is often used in the literature), here  we will keep again the shift-invariance explicit since, in our parametrization, such choice would simplify intermediate expressions only marginally.

We will next show that, out of the eight initial parameters, only three have an impact on
the \emph{shape} features of the potentials with two degenerate minima. Two parameters can be removed thanks to the
shift-symmetry and the requirement of degeneracy of the minima. Two more
parameters can be removed by rescaling the potential in the $s$ and
$h$ directions. Finally the overall scale of the potential has no qualitative meaning in this discussion, leaving us with three parameters. In the
following we present a parameter choice, a refinement with respect to eqs.~(\ref{reli})-(\ref{relf}), which is especially handy in
describing the qualitative features of a potential with degenerate minima.

To discuss the minima, let us consider the curves $\partial V/\partial
h=0$ and $\partial V/\partial s=0$ more systematically.  The curves
$h=0$ and $D^2_h(s)$, at which $\partial V/\partial h=0$, were already
introduced in eqs.~(\ref{h0h1}). We begin with $\lambda_m\neq 0$, in which case we can rewrite the curve $D_h^2(s)$
as:
\be
\label{eq:def_barh}
D_h^2(s) = \bar h^2 - \frac{\lambda_m v^2}{m_h^2} (s-w_p)^2\ ,
\ee
where, using both the notation of eqs.~(\ref{reli})-(\ref{relf}) and the original notation,
\bea
\bar h^2 &\equiv&v^2 + \frac{ m_{sh}^4}{\lambda_m m_h^2}=\frac{1}{\lambda_h}\left(\mu_h^2+\frac{\mu_m^2}{8\lambda_m}\right)
  \ ,\\
w_p &\equiv& w - \frac{m_{sh}^2}{\lambda_m v}=-\frac{\mu_m}{2\lambda_m}  \ .
\eea
In the plots, we will generally choose our singlet coordinates to have $w_p=0$ (setting $\mu_m=0$), so that the parabola $D_h^2(s)$ has its axis at $s=0$. As
discussed in Section~\ref{sec:PhT}, the minimum at $h=0$ can only be
located in the regions with $D_h^2(w_0)<0$.

\begin{figure}[t]
\includegraphics[width=0.95\textwidth, clip ]{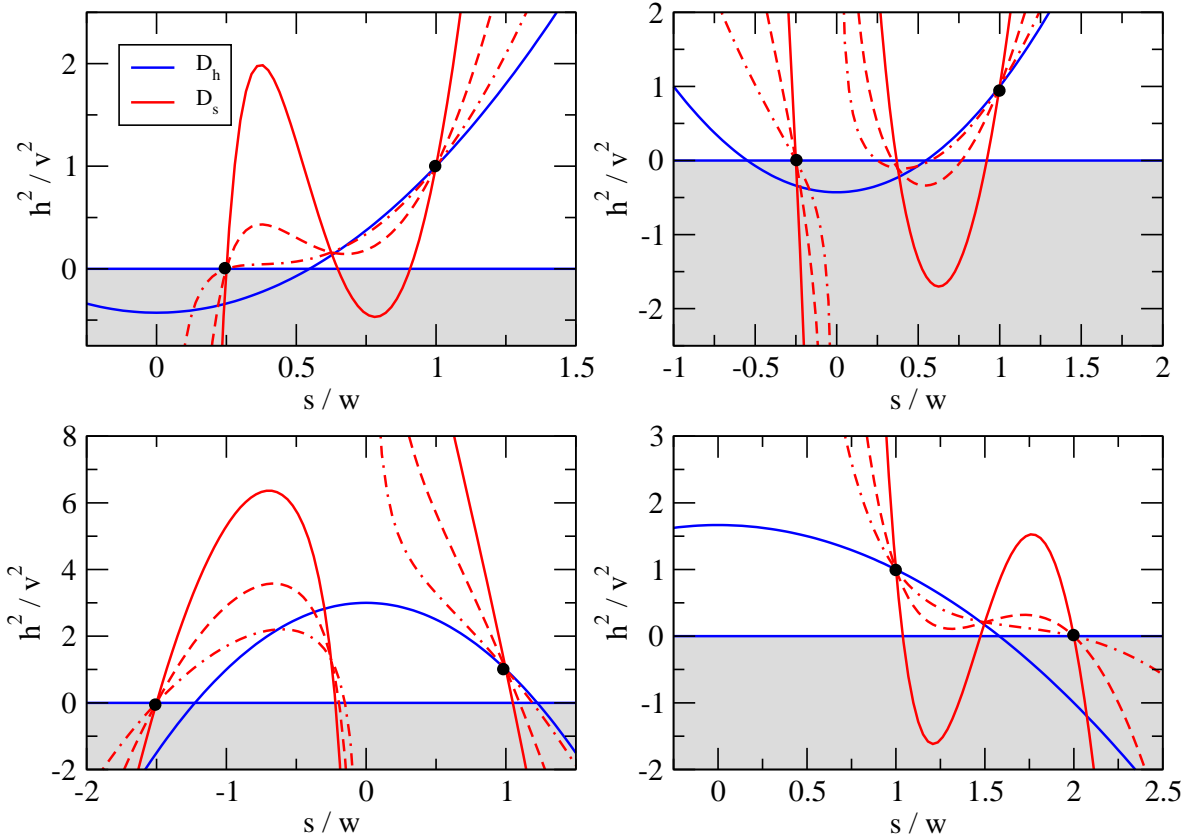}
\caption{\label{fig:cases}
Curves $D_h^2(s)$ (solid blue) and several $D_s^2(s)$ (red solid, dashed and dash-dotted) with different values of $m_s^2$, intersecting to give two potential minima (indicated by black dots) in the 4 different cases listed in eqs.~(\ref{casea})-(\ref{cased}).}
\end{figure}

The potential is a quartic in $s$ and hence has at most three
extrema with respect to $s$ for
fixed $h$ (out of which at most two are minima). For general values of $h$,
\be
\frac{\partial V}{\partial s} = \frac14 h^2 (\mu_m + 2 \lambda_m s)
+ (\mu_1^3 + \mu_s^2 s+ \mu_3 s^2  + \lambda_s s^3 ),
\ee
and the curve $\partial V/\partial s=0$ in the $(h^2, s)$-plane as a function of
$s$ is given by
\be
\label{h22}
h^2=D_s^2(s) \equiv -4\frac{\mu_1^3+ \mu_s^2 s + \mu_3 s^2 + \lambda_s s^3 }
{\mu_m + 2\lambda_m s }\ ,
\ee
or, in our parametrization:
\be
D_s^2(s) = v^2 -\frac{(s-w)}{\lambda_m m_h^2(s-w_p)}\left[
2m_h^2m_s^2+3v(\lambda_m m_{sh}^2+ 4m_*v)(s-w)+(4\lambda^2+\lambda_m^2)(s-w)^2\right]\ .
\ee
This function is single-valued for fixed $s$ and has a pole at $s=w_p$. When
this pole is canceled by a zero of the numerator, the line $s=w_p$
is also a solution of $\partial V/\partial s=0$  (this is e.g.~the case
for a potential with a $\mathbf{Z}_2$ symmetry $s \to -s$) and enters the discussion.  The asymptotic
behavior of $D_s^2(s)$ at large $s$ is $D_s^2(s) \to - 2\lambda_s
s^2/\lambda_m$ and hence is qualitatively different depending on the sign of
$\lambda_m$. In the following, we distinguish four
different cases given by\footnote{The case $w_0=w_p$ is only possible for $\lambda_m<0$ and smoothly connects cases $(a)$ and $(b)$. This case will be very relevant in the particular scenario of Section~\ref{sec:Z2}. The case $w=w_p$, on the other hand, is only possible for $\lambda_m>0$ and smoothly connects cases $(c)$ and $(d)$.}
\bea
\label{casea}
(a) && \, \lambda_m<0\ ,\;\;\;\; (w_0 - w_p) (w - w_p) >0\ ,  \\ 
(b) && \, \lambda_m<0\ ,\;\;\;\; (w_0 - w_p) (w - w_p) <0\ , \\
(c) && \, \lambda_m>0\ ,\;\;\;\; (w_0 - w_p) (w - w_p) <0\ ,\\
(d) && \, \lambda_m>0\ ,\;\;\;\; (w_0 - w_p) (w - w_p) >0\ .\label{cased}
\eea
The sign of $\lambda_m$ determines whether the parabola $D_h^2(s)$
curves up or down.  The sign of $(w_0 - w_p) (w - w_p)$ determines if
the two minima lie at the same side or different sides of the pole at
$s=w_p$. Examples for the curves $D^2_s(s)$ and $D^2_h(s)$ in
these four cases are depicted in Fig.~\ref{fig:cases}.
Is it obvious
that, once we have two degenerate minima in our potential,
there is a barrier separating them. The two minima and the saddle
point in between appear at the intersections between the $D^2_h(s)$ and
$D^2_s(s)$ curves, which must be of sufficiently high degree to allow
for such  structure.\footnote{Some analysis in the literature solve the equation $\partial V/\partial s=0$ for $s(h)$ and then look for a barrier in the one-dimensional potential $V[h(s),s]$. While this is justified in some cases, Fig.~\ref{fig:cases} illustrates some of the possible dangers of this procedure: $h(s)$ might not be single-valued; $V[h(s),s]$ can venture into the unphysical region; and sometimes there is no continuous path connecting both minima and having $\partial V/\partial s=0$.}
The different cases listed above will in general lead to different shapes of the potential barriers, which  have an impact on the profile of the critical bubbles for the EWPhT.

It is convenient to introduce a different parametrization for this degenerate case, that can be easily connected  to
the qualitative features of the curves $D^2_h(s)$ and $D^2_s(s)$. Remember that
the solutions of $\partial V/\partial s=0$ at the axis $h=0$
lead to a cubic equation and up to two local minima that are
cumbersome to determine analytically. To avoid this problem, it is
helpful to treat the position of the minimum $w_0$ as a free parameter
and trade it for the parameter $m_*$. Imposing the condition
that both minima are degenerate, the parameter $\lambda^2$ can be
fixed. Finally, one can also trade the parameter $m^2_{sh}$ for $w_p$, the 
point that marks the axis of symmetry of the curve $D^2_h(s)$. We end up with the following parameters
\be
\{ w, w_p, w_0, v, m_h^2, m_s^2, \lambda_m\}\ ,
\ee
related to the ones in eq.~(\ref{eq:params1}) by
\bea
\label{mstardeg}
m_* &=& \frac{\Delta w}{4}\left\{-\lambda_m \left[\frac{m_h^2}{\Delta w^2}+
\frac{m_{sh}^2}{v\Delta w}\right]
+\frac{m_h^2}{\Delta w^2}\left[\frac{m_h^2}{\Delta w^2}+
2\frac{m_s^2}{v^2}+3\frac{m_{sh}^2}{v\Delta w}\right]
\right\}\ , \\
\label{lambda2deg}
\lambda^2 &=& \lambda_d^2\equiv-\frac{1}{4}\left[\lambda_m+\frac{m_h^2}{\Delta w^2}\right]^2
+\frac{m_h^2}{\Delta w^2}\left[\frac{m_h^2}{\Delta w^2}+\frac{m_s^2}{v^2}+2\frac{m_{sh}^2}{v\Delta w}\right]\ ,\\
m_{sh}^2 &=& \lambda_m v (w - w_p)\ , 
\eea
where\footnote{$\Delta w$ cannot be zero in the presence of a barrier, and therefore, at the EWPhT there is a jump both in $\langle h\rangle$ and  $\langle s\rangle$. This can be important for some EW baryogenesis mechanisms.} $\Delta w\equiv w-w_0$ and, as promised, we give the expression
for $\lambda_d^2$, the value of $\lambda^2$ required for degeneracy of the minima (and already discussed at the end of SubSection~\ref{localminima0}). 
Some of the particular combinations of masses that appear above have a direct
physical interpretation. For instance, we have
\be
\frac{m_h^2}{\Delta w^2}+\frac{m_s^2}{v^2}+2\frac{m_{sh}^2}{v\Delta w}=
\left(\frac{1}{v^2}+\frac{1}{\Delta w^2}\right)m_\varphi^2\ ,
\ee
where $m_\varphi^2$ is the squared mass at the broken minimum along the 
direction $\varphi\equiv s \cos\theta  + h \sin\theta $ (that joins both minima),
where the angle $\theta$ satisfies $\tan\theta=v/\Delta w$.

Finally, we can now obtain the condition necessary to ensure that $w_0$ is the deepest
minimum along $h=0$, which simply reads:
\be\label{condDET}
{\rm Det}{\cal M}_s^2 > \frac{v^2}{\Delta w^2}(m_h^2|_0)m_h^2\ ,
\ee
where $m_h^2|_0$ is the squared-mass $\partial^2 V/(\partial h)^2$ at the symmetric minimum
$(0,w_0)$, which is given by
\be\label{CondH}
m_h^2|_0=\frac{\Delta w^2}{2}\left[\lambda_m-\frac{m_h^2}{\Delta w^2}-2\frac{m_{sh}^2}{v\Delta w}\right]>0\ .
\ee
It is interesting that these conditions also ensure that the potential is bounded from below: that is, if (\ref{condDET}) and (\ref{CondH}) are satisfied then $\lambda^2$ fulfills the stability constraints discussed in SubSection~\ref{stability}.

The fact that the quantities in the discussion above depend always on the ratios $R_h\equiv m_h^2/\Delta w^2$, 
$R_s\equiv m_s^2/v^2$ and $R_{sh}\equiv m_{sh}^2/(v\Delta w)$ (plus $\lambda_m$) is a consequence 
of the fact that, as we mentioned  earlier, the conditions for degeneracy are independent of possible rescalings of the fields $h$ and $s$. If such field rescaling is followed by a global rescaling of the potential so as to leave $\lambda_m$ unchanged, the above ratios are also invariant under such combination of rescalings. Multiplying these 4 parameters by a common factor changes the potential by the same overall factor and so, the shape of the potential with two degenerate minima is determined by just 3 shape parameters: the three independent ratios  $\{\lambda_m/R_{h},R_{sh}/R_h,R_s/R_h\}$.

The potential with such degenerate vacua takes the  form
\bea
V & = & \frac18 v^2 m_h^2 \left\{(h_r^2-1)^2
+(s_r-1)^3 (1+3s_r)+ 2 \frac{\lambda_m}{R_h} (s_r-1)^2
(h_r^2 -s_r^2)\right.\nonumber\\
&&\hspace{1.5cm}+\left.4\frac{R_s}{R_h}s_r^2(s_r-1)^2
+4\frac{R_{sh}}{R_h}(s_r-1)[h_r^2+s_r^2(2s_r-3)]
\right\}\ ,
\label{Vdgen}
\eea
where we have used
\be
h_r\equiv \frac{h}{v}\ ,\quad
s_r\equiv \frac{s-w_0}{\Delta w}\ ,
\ee
with the EW minimum at $h_r=1, s_r=1$ and the symmetric one at $h_r=s_r=0$.

The previous discussion assumes $\lambda_m\neq 0$, but
one can also get degenerate minima for $\lambda_m=0$. In that case, the curve $D^2_h(s)$ becomes a straight line and reads
\be
h^2=D^2_h(s)=v^2  - 2 v (s-w)\frac{m_{sh}^2}{m_h^2} \ ,
\ee
intersecting the axis $h=0$ at 
\be
\label{wx}
w_x =w+ \frac{m_h^2 v}{2 m_{sh}^2}\ .
\ee
We can again impose degeneracy of the two vacua, and write $m_*$ in terms of $w_0$, as we did for $\lambda_m\neq0$, arriving at the same eqs.~(\ref{mstardeg}) and (\ref{lambda2deg}); on the other hand, $m_{sh}^2$ is  now fixed by (\ref{wx}):
\begin{equation}
m_{sh}^2=\frac{m_h^2 v}{2(w_x-w)}\ .
\end{equation}
The parameters to describe the potential with degenerate minima in the case $\lambda_m=0$ are then
\be
\{ w, w_x, w_0, v, m_h^2, m_s^2 \}\ .
\ee

\subsubsection{\em Flat Directions}

An intriguing situation occurs in the particular limiting case in which 
$D^2_{h,s}(s)$ fall on top of each other; then these curves 
correspond to a flat direction of the potential. In terms of the original potential parameters, $D_h^2(s)\equiv D_s^2(s)$ requires 
\bea
\mu_h^2\mu_m+4\lambda_h \mu_1^3 &=&0\ ,\\
4\lambda_m \mu_h^2-\mu_m^2+8\lambda_h \mu_s^2&=&0\ ,\\
8\lambda_h \mu_3-3\lambda_m\mu_m&=&0\ ,\\
4\lambda_h\lambda_s-\lambda_m^2&=&0\ ,
\eea
which, in terms of our parameters, simply read\footnote{Here there are only three conditions, since the shift symmetry has not been used to fix one of the parameters in the original parametrization. In (\ref{robot}), on the other hand, this degree of freedom disappears since the parameters are shift-independent.}
\be\label{robot}
\lambda^2=0\ ,\quad m_*=0\ ,\quad {\rm Det}{\cal M}_s^2=0 \ .
\ee
While these conditions would be unacceptable at
$T=0$, we will see in later Sections that, if such flat direction develops at the critical temperature for the electroweak phase transition, the strength of this transition 
can be significantly larger: the small effect from thermally induced cubic terms in the finite-temperature potential is enhanced along such flat directions.

Imposing the conditions derived above, the potential takes the simple form
\be
V=\frac{1}{8m_h^2 v^2}\left[m_h^2(h^2-v^2)+\lambda_m v^2(s-w)^2
\pm 2m_hm_s v (s-w)
\right]^2\ ,
\ee
where the $\pm$ sign corresponds to the possible sign of $m_{sh}^2$.
The flat direction will in general be parabolic in the 
$(h^2,s)$-plane, becoming a straight line for $\lambda_m=0$. 
 We will examine this scenario in more detail in the following Sections.

\section{Strong Electroweak Phase Transitions\label{sec:strong_PT}}

The present model can develop very strong phase transitions
if the barrier separating the broken and symmetric vacua is produced
by tree-level effects (as discussed in the previous Section) and not by the cubic 
term resulting from one-loop thermal contributions of bosons (which is the most studied
mechanism to achieve a first-order phase transition). Indeed, the latter are always proportional to the temperature and lead to a critical order parameter $v_c\propto T_c$. In this case the strength of the EWPhT, characterized by $v_c/T_c$, is independent of the temperature and typically proportional to small couplings. For a tree-level barrier, on the other hand, $v_c$ is proportional to other dimensionful parameters of the potential and almost T-independent\footnote{Incidentally, due to this property the strength of such transitions will be insensitive to the gauge-fixing subtleties  discussed in \cite{Gauge}.}. In this case $v_c/T_c$ can be large for small critical temperatures.

\subsection{Evolution of Parameters at Finite $T$}

When the barrier is produced at tree-level, it is enough to include in the
one-loop thermal potential the leading terms in the high-temperature
expansion that lift the minimum in the broken phase. In our model, these terms are
\be
\label{eq:1loop_highT}
V_{1-loop}^{T\neq0}=  \left(\frac{1}{2}c_h h^2 + \frac{1}{2}c_s s^2 + m_3 s\right)T^2\ ,
\ee
where
\bea
c_h&=&\frac{1}{48}\left[9g^2+3{g'}^2+2(6h_t^2+12\lambda_h+\lambda_m)\right]\ ,\nonumber\\ 
c_s&=&\frac{1}{12}(2\lambda_m+3\lambda_s)\ , \nonumber\\
m_3&=&\frac{1}{12}(\mu_3+\mu_m)\ .  
\label{thermalcs}
\eea
Here $g$ and $g'$ are the $SU(2)_L$ and $U(1)_Y$ gauge couplings and
$h_t$ is the top Yukawa coupling. Additional particles coupled to the Higgs or the singlet will in general contribute to these quantities.  At very high temperature the potential is dominated by this contribution, which
drives $\langle h\rangle\rightarrow 0$, restoring the EW symmetry \cite{Linde}. On the other hand,
the singlet develops a thermal tadpole so that $\langle s\rangle\rightarrow s_\infty=
-m_3/c_s$ at high $T$.\footnote{\label{foot:shiftm3}Under a singlet shift, $m_3\rightarrow m_3+c_s\sigma$, so that $s_\infty$ transforms as it should.} In the general case, without an $s\rightarrow -s$ symmetry,
there is no reason to expect $s\rightarrow 0$. This could be arranged by using
the coordinate frame $\mu_3=-\mu_m$ (provided $c_s>0$) but there is no
sense in which a symmetry associated with the singlet is being
restored, simply because there is no symmetry.\footnote{In the case with a $\mathbf{Z}_2$-symmetric potential
thermal corrections do not  break the symmetry and
$s\rightarrow 0$ at high $T$, restoring the symmetry in the vacuum.}  

The key point 
in our approach is that the terms in (\ref{eq:1loop_highT}) can be absorbed in the definition
of $T$-dependent parameters 
\bea
\label{massT}
-\mu_h^2(T)&\equiv&-\mu_h^2+c_h (T^2-T_c^2)\ ,\nonumber\\
\mu_s^2(T)&\equiv&\mu_s^2+c_s (T^2-T_c^2)\ ,\\
\mu_1^3(T)&\equiv&\mu_1^3+m_3 (T^2-T_c^2)\ ,\nonumber
\eea
where we use a notation in which, when no temperature is indicated for some
$T$-dependent quantity, it is implicitly assumed that its value at $T_c$ is meant, {\em e.g.} $m_s^2\equiv m_s^2(T_c)$.
We can then apply the general results on the structure of the potential
derived in Section~\ref{sec:PhT}.  The minima are still determined by
the curves $D_h^2(s)$ and $D_s^2(s)$, which are now $T$-dependent and induce a $T$-dependence in the location of the minima. We have
\be
\label{DT}
\frac{dD_h^2(s)}{dT^2}=-\frac{c_h}{\lambda_h}\ ,\quad
\frac{dD_s^2(s)}{dT^2}=-\frac{4(m_3+ c_s s)}{\mu_m+2\lambda_m s}\ .
\ee
The curve $D_h^2(s)$  approaches the axis $h=0$ as $T$ increases (keeping fixed its symmetry axis and without changing its shape); this guarantees that $v\rightarrow 0$ at high $T$. The evolution of $D_s^2(s)$ is more complicated in general. We can simplify somewhat the analysis by choosing $\mu_m=0$ through the shift-symmetry, and then we have
\be
\frac{dD_s^2(s)}{dT^2}=-\frac{2}{\lambda_m}\left(c_s-\frac{m_3}{s}\right)\ ,\quad (\mu_m=0)\ .
\ee

Now, for the cosmological history from $T_c$ to $T=0$ to be acceptable, the 
EW minimum must be the global one at $T=0$: $V_b(0)<V_s(0)$. This
requirement will put a constraint on the parameters of the potential. 
The evolution of the difference $\Delta V_{bs}(T)\equiv V_{b}(T)-V_{s}(T)$ with $T$ can be determined as follows:
\be
\frac{d\Delta V_{bs}(T)}{dT^2}=\left.\sum_i\left[\frac{\partial V}{\partial\mu_i^2}\right|_b-\left.\frac{\partial V}{\partial\mu_i^2}\right|_s\right]\frac{d\mu_i^2}{dT^2}+\left.\sum_\alpha\left[\frac{\partial V}{\partial\phi_\alpha}\right|_b\frac{d\langle\phi_\alpha\rangle_b}{dT^2}-\left.\frac{\partial V}{\partial\phi_\alpha}\right|_s\frac{d\langle\phi_\alpha\rangle_s}{dT^2}\right]\ ,
\ee
where we symbolically write $\mu_i^2\equiv\{\mu_h^2,\mu_s^2,\mu_1^3\}$, $\phi_\alpha\equiv\{h,s\}$. Noting that $\partial V/\partial\phi_\alpha=0$ at both minima and using the $T$ dependence of the $\mu_i^2$ parameters from eq.~(\ref{massT}), we obtain
\be
\label{deepcond}
\frac{d\Delta V_{bs}(T)}{dT^2}=\frac{1}{2}\left\{c_h v^2(T)+
\Delta w(T)\left(c_s[w(T)+w_0(T)]+2m_3\right)\right\}\ ,
\ee
(which is a shift-invariant expression, see footnote~\ref{foot:shiftm3}). A necessary condition for the EWPhT to take place  is that
this derivative is positive at $T_c$ so that the broken minimum
is the deepest one at least for $T\simlt T_c$, 
\be
\label{deepcond2}
\left.\frac{d\Delta V_{bs}(T)}{dT^2}\right |_{T_c}=\frac{1}{2}\left\{c_h v^2+
\Delta w\left(c_s[w+w_0]+2m_3\right)\right\}>0\ .
\ee
 Note, however, that this is a necessary but not sufficient condition to guarantee that $(v,w)$ is the global minimum at $T=0$: this must be checked separately, as summarized in 
Table~\ref{Table1}, and as will be shown in particular examples in later Sections.

\subsection{Strategy to Identify Strong EWPhTs}

What are the regions of parameter-space that lead to a tree-level barrier? In terms of the original parameters of the potential in 
eq.~(\ref{Vtree}), the answer to this question generally involves a complicated superposition of non-linear conditions, with hidden physical meaning and hard to use for phenomenological applications. This task is greatly simplified by the parametrization introduced in Section~\ref{sec:PhT}, which allows an easy identification of a potential with stable minima. Moreover, when minima exists they must necessarily be separated by a barrier.  Indeed, any potential of the general form (\ref{newpot}), has a stable global minimum at $(v,w)$ for any values of the parameters $\{ v, w, m_{h}^2, m_{s}^2, m_{sh}^2, \lambda_m, \lambda^2, m_*\}$ if the simple conditions discussed below eqs.~(\ref{eq:detM}),(\ref{V444}) and (\ref{lambdaPot2min}) are fulfilled. Similarly, a potential of the degenerate form (\ref{Vdgen}), with parameters $\{ w, w_p, w_0, v, m_h^2, m_s^2, \lambda_m\}$ satisfying the conditions of eqs.~(\ref{condDET})-(\ref{CondH}), 
has a barrier between two degenerate minima at $(v,w)$ and $(0,w_0)$. Both cases are summarized in Table~\ref{Table1}.
\begin{table}[tdp]
\begin{center}
\begin{tabular}{|c||c|l|}
\hline
& Parameters &Conditions\\
\hline
\hline
$\begin{array}{c} T = T_c\\ \textrm{Degenerate $V$, (\ref{Vdgen})}\end{array}$ &$ \begin{array}{l}\{ w, w_p, w_0,\\ v, m_h^2, m_s^2, \lambda_m\}\end{array}$ &$ \begin{array}{l} \textrm{Stability in $w_0$ and $w$, (\ref{condDET})-(\ref{CondH})}: \vspace{2mm}\\ {\rm Det}{\cal M}_s^2 > (v^2/\Delta w^2)(m_h^2|_0)m_h^2 \\ m_h^2|_0,m_h^2,m_s^2>0 \end{array}$\\
\hline
Matching &$ \begin{array}{l} \lambda^2=\lambda^2_d\\m_*=m_*(w_0)\\m^2_{sh}=m^2_{sh}(w_p) \end{array}$& $\begin{array}{l}\textrm{Broken min. deepest, (\ref{deepcond2})}:\vspace{2mm}\\ \left.d\Delta V_{bs}(T)/dT^2\right |_{T_c}>0\end{array}$\\
\hline
$\begin{array}{c} T \leq T_c\\ \textrm{General $V$, (\ref{newpot})}\end{array}$ &$\begin{array}{l}\{ v, w, m_{h}^2, m_{s}^2,\\ m_{sh}^2, \lambda_m, \lambda^2, m_*\}\end{array}$ & $\begin{array}{l} \textrm{$V$ bounded below, (\ref{V444})}:\vspace{2mm}\\ \lambda^2>0 \quad \quad \,\, (\lambda_m\leq 0)\\  \lambda^2>-\lambda_m^2/4 \quad (\lambda_m> 0)\vspace{3mm} \\  \textrm{Vacuum stability, (\ref{eq:detM})}:  \vspace{2mm} \\ \mathrm{Det}{\cal{M}}^2_s>0 \\ m_h^2,m_s^2>0\vspace{3mm} \\ \textrm{Global min., (\ref{lambdaPot2min})}:\vspace{2mm}\\  \lambda^2 \geq 8\tilde{\lambda}^2/9 \end{array}$\\
\hline
\end{tabular}
\end{center}
\caption{Summary on the strategy and parameter conditions to identify potentials with large tree-level barriers.\label{Table1}}
\end{table}

Hence, the strategy to find a model with a strong phase transition is the following, as illustrated in Table~\ref{Table1}. First, 
choose a value for  the parameters $\{ w, w_p, w_0, v, m_h^2, m_s^2, \lambda_m\}$, subject to the simple conditions summarized in the upper part of Table~\ref{Table1}. Any such choice determines a potential of the form (\ref{Vdgen}) with two degenerate minima with broken and unbroken EW symmetry
and a barrier separating them: this will be the thermal potential at some critical temperature $T_c$,
which at this point we are free to choose. Once we select $T_c$, we can match this potential to a general potential (\ref{newpot}), making sure to satisfy the conditions in the middle part of Table~\ref{Table1}, which ensure that the broken minimum gets deeper than the symmetric minimum for decreasing $T\lesssim T_c$. Finally, using the formulae outlined in the previous Subsection, we can evolve all
the parameters with $T$ to obtain their values at $T=0$: these are the relevant parameters that enter physically meaningful quantities like the scalar masses, mixings, etc. As $T$ is lowered, it is crucial that our (broken) vacuum remains the global stable minimum of the potential: this is guaranteed by the conditions in the lower part of Table~\ref{Table1}. In fact, the stability conditions on $\lambda^2$ are guaranteed to be satisfied
once the $T_c$ parameters satisfy the conditions in the upper part of Table~\ref{Table1}. Eventually, $h$ and $s$ should be suitably rescaled to ensure that the zero temperature vev is
$v=v_{EW}$. In this way, different values of $T_c$ will generate a family of models with different values of the potential parameters, but all with the same potential shape at $T_c$. 

With the zero temperature potential at hand, a full-fledged one-loop analysis can be performed to confirm the first-order nature of the phase transition and to calculate the real critical temperature at which the broken and symmetric vacua are degenerate. This temperature will in general differ from the $T_c$ parameter we have used, which corresponds in the mean-field approximation to the real critical one. We call $T_{c,MF}$ the latter and $T_{c,1L}$ the former.  We will show this strategy at work in some examples in later Sections. The details of the calculation of the one-loop scalar potential at finite $T$, which are standard, are relegated to Appendix~\ref{app:thermalpot}. 

Notice that our estimate of $v_c/T_c$ is conservative since the true critical temperature at which the transition starts (the nucleation temperature) is smaller than our $T_c$. For a complete analysis, one should also recalculate how the sphaleron energy is affected by the Higgs barrier and how this impacts the critical ratio $v_c/T_c$ required for a successful preservation of the baryon asymmetry. However, it is generically the case that the sphaleron energy is dominated by gauge degrees of freedom with Higgs effects amounting to a few percent change 
(see {\em e.g.} \cite{sphaleron}).

Before moving to the examples, let us finally mention the case in which the potential 
has two degenerate minima but both with $h>0$. Does this correspond to some situation
of physical interest? We know that, also in such cases, the EW symmetry will be restored at
some higher temperature and a local minimum at $h=0$ will arise, at which point there can be only one broken minimum (as we showed in the previous Section). Hence, in this
case either the two broken minima merge together or one of them moves to $h=0$ before the critical temperature is reached. Both options correspond to peculiar phase transition histories
and merit study, which we leave for a future analysis.

\section{Special Cases: $\mathbf{Z}_2$-symmetric Potential\label{sec:Z2} }  

In a general study like this one, concentrating on the potential (\ref{Vtree}) which involves many parameters, it is crucial to identify whether or not some regions of parameter space are more natural than others. This point is especially relevant in the presence of symmetries, which select a region of parameter space with vanishing volume (and hence unlikely from the point of view of a general analysis) and preserve it under RG-evolution and at finite temperature. For the SM plus a singlet, the only symmetry (both of the kinetic terms and of the potential at renormalizable level) that is interesting from the point of view of the EWPhT is the $\mathbf{Z}_2$ symmetry $s\rightarrow -s$.
One particular case of interest that falls in this category is the
so-called Singlet Majoron Model \cite{Majoron}. The EWPhT in this model has been studied in \cite{EWPHTMajoron}.\footnote{Although in the model of \cite{EWPHTMajoron} the scalar is complex (it carries lepton number), from the point of view of the potential for the real part of $s$, it reduces to our case.}

Making the $\mathbf{Z}_2$ symmetry manifest (although the general analysis of Section~\ref{sec:PhT} can be carried out without problem), the
potential is of the form given in eq.~(\ref{Vtree}) with 
\be 
\label{eq:Z2conditions}
\mu_1=0\
,\;\;\;\; \mu_m = 0\ , \;\;\;\; \mu_3 = 0\ .  
\ee 
In terms of our  parameters (\ref{eq:params1}), these constraints  translate into two separate branches, depending on whether the $\mathbf{Z}_2$-symmetry is broken spontaneously or not. The $\mathbf{Z}_2$-symmetric case has
\begin{equation}
w=0\ ,\quad m_*=0\ ,\quad m_{sh}^2=0\ ,
\end{equation}
and the $\mathbf{Z}_2$-broken case has $w\neq 0$ and
\be
\label{Z2blight}
m_s^2=2\lambda_s w^2=(4\lambda^2+\lambda_m^2)\frac{v^2w^2}{m_h^2}
\ ,\quad m_{sh}^2=\lambda_m v w\ ,\quad m_*=\lambda^2 w\ ,
\ee
which allows to extract the usual parameters $w$, $m_*$ and $m_{sh}^2$ in terms of the others:
\begin{equation}
w=\frac{m_hm_s}{v\sqrt{4\lambda^2+\lambda_m^2}}\ ,\quad m_*=\lambda^2\frac{m_hm_s}{v\sqrt{4\lambda^2+\lambda_m^2}}\ ,\quad m_{sh}^2=\lambda_m\frac{m_hm_s}{\sqrt{4\lambda^2+\lambda_m^2}}\ .
\end{equation}
This model can then be described by the 5 parameters $\{v,m_h^2,m_s^2,\lambda_m,\lambda^2\}$ in both branches but, to avoid confusion, the first part of this Section will be clearer in the standard notation of eq.~(\ref{Vtree}).

At high temperature we expect the minimum to lie at the symmetric
point $h=0$, $s=0$ but it might happen that $s\neq 0$ prior to the
EWPhT. The stationary points of the potential will be determined by the intersections of the curves $\partial V/\partial h=0$ and $\partial V/\partial s=0$, which now have very simple expressions:
\bea
\frac{\partial V}{\partial h}=0 &\Rightarrow &\left\{h=0\ ,\;\;\;\;
 {\rm and } \;\;\;\;
h^2=D^2_h(s)=\frac{1}{2\lambda_h}(2\mu_h^2-\lambda_m s^2)\right\}\ ,\nonumber\\
\frac{\partial V}{\partial s}=0 &\Rightarrow &\left\{s=0\ ,\;\;\;\;
 {\rm and } \;\;\;\;
h^2=D^2_s(s)=-\frac{2}{\lambda_m}(\mu_s^2+\lambda_s s^2)\right\}\ .
\label{parabolas}
\eea
That is, now $D_h^2(s)$ and $D_s^2(s)$ are just parabolas with the same axis of symmetry, at $s=0$, and different widths\footnote{Which parabola is widest depends on the relative size of the two widths $\lambda_m/(2\lambda_h)$ vs. $2\lambda_s/\lambda_m$ and is therefore controlled by the sign of $\lambda^2$.}.
Furthermore, to understand the nature of stationary points along the branch $s=0$, it is useful 
to write
\be
\label{massa}
\left.\frac{\partial^2V}{(\partial s)^2}\right|_{s=0}=
\frac{1}{2}\lambda_m[h^2-{\bar h}_s^2]\ .
\ee
meaning that, for $\lambda_m>0$ ($\lambda_m<0$) minima along $s=0$ can only appear for $h^2$ above (below) 
the vertex ${\bar h}_s^2=-2\mu_s^2/\lambda_m$ of the $D_s^2(s)$ parabola.

\begin{figure}[t]
\includegraphics[width=0.45\textwidth, clip ]{figs/Ds_Z2.eps}
\hspace*{0.1cm}
\includegraphics[width=0.6\textwidth, clip ]{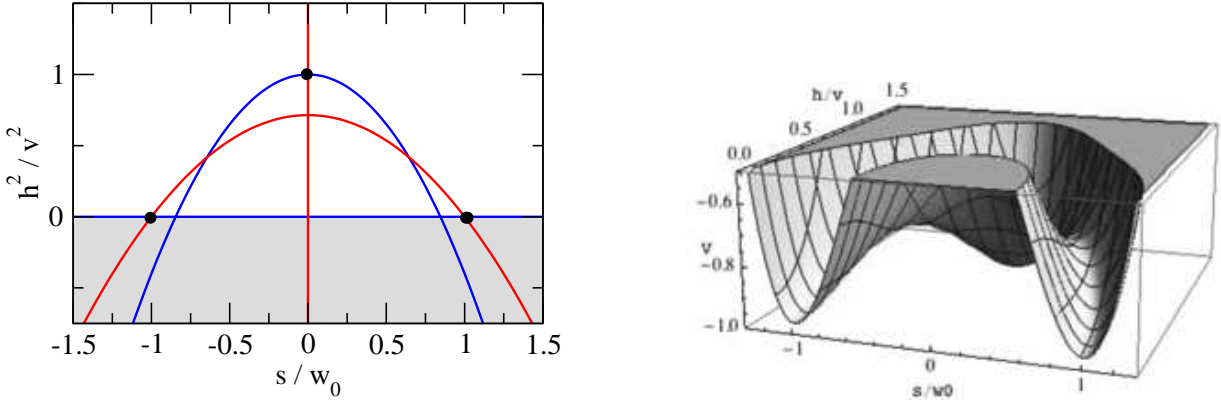}
\caption{\label{fig:Z2case}
Special scenario with $\mathbf{Z}_2$ symmetry, $\lambda_m>0$ and $\lambda^2<0$. 
Left, curves with $\partial V/\partial h=0$ [$D^2_h(s)$ and $h=0$, blue lines] and $\partial V/\partial s=0$ [$D^2_s(s)$ and $s=0$, red lines] intersecting in the minima at $(0,\pm w_0)$
and $(v,0)$, as indicated by the black dots. Right, corresponding potential showing the barrier between minima.}
\end{figure}

Can this  constrained setting give rise to a tree-level barrier?
As we saw in the previous Section, in order to arrange for two degenerate minima, 
one at $h=0$ and the other at $h\neq 0$, no tree-level barrier can appear if the broken minimum has $w\neq 0$. 
This is because the $\mathbf{Z}_2$-symmetry enforces the existence of two minima with $s=\pm w$ but this prohibits a minimum at $h=0$,
as discussed below eq.~(\ref{mass0}).\footnote{In terms of the two parabolas $D^2_{h,s}(s)$, it is difficult to arrange that they cut twice,  at the broken minimum and at the saddle point in between, because they  have the same axis of symmetry, at $s=0$.} Hence, the $\mathbf{Z}_2$-symmetric case can only have a tree level barrier at the critical
temperature if the minimum with $h^2>0$ lies at the symmetry axis
$w=w_p=0$. This situation is illustrated by Fig.~\ref{fig:Z2case} which shows the intersecting
curves $D^2_{h,s}(s)$ in the $(h^2/v^2,s/w_0)$-plane (left plot) and the corresponding potential with its barrier (right plot). We focus on this particular case in the rest of this Section.

Following the approach of Section~\ref{sec:PhT}, we identify the conditions for a barrier separating the broken and unbroken minima. From eq.~(\ref{mass0}), we know that along $h=0$, stable minima $(0,\pm w_0)$ require $D_h^2(w_0)<0$ which leads to
\be\label{lambdadegcondition}
\lambda^2<-\frac{\lambda_m m_s^2}{2v^2}<0\ ,\quad \lambda_m>0\ ,
\ee
where $w_0$, solution of  $D_s^2(w_0)=0$, is given by
\be
w_0^2=-\frac{\mu_s^2}{\lambda_s}=\frac{m_h^2(\lambda_m v^2-2m_s^2)}{v^2 \left(4 \lambda ^2+\lambda_m^2\right)}\ .
\ee
Furthermore, the condition of degeneracy, eq.~(\ref{lambda2deg}), imposes
\be
\lambda^2=\frac{m_s^4-\lambda_m v^2m_s^2}{v^4}\ .
\label{degcond}
\ee
This condition can be rewritten  as
\be
\label{degconds}
m_s^2=\frac{v^2}{2}\left(\lambda_m-2\sqrt{\lambda_h\lambda_s}\right)\ ,
\ee
which will be useful later on.

In this degenerate case, the potential then takes the form
\be
\label{VcZ2}
V = \frac18 m_h^2 v^2 \left[ 
4 \frac{R_s}{R_h} \frac{h^2 s^2}{v^2 w_0^2}
+ \left( \frac{h^2}{v^2} + \frac{s^2}{w_0^2} - 1 \right)^2
\right]\ ,
\ee
showing a concrete example in which the overall shape of the potential is controlled by the ratio $R_s/R_h$. This is now
the only relevant shape parameter (as was to be expected starting with only 5 d.o.f.s and removing 3 for rescalings, 1 for degeneracy and no shift freedom) and it controls the height of
the barrier that separates the symmetric and broken minima. For comparison with other cases, notice that this degenerate $\mathbf{Z}_2$ scenario corresponds in fact to the shape parameters  $\lambda_m/R_h=1+2R_s/R_h$
and $R_{sh}/R_h=0$.

\subsubsection{\em Case with Flat Directions}

Applying the general discussion of flat directions in
 Section~\ref{sec:PhT} to the particular case of the $\mathbf{Z}_2$-symmetric scenario, we see that a flat direction arises  for
\be
\frac{\mu_h^2}{\lambda_h}=-\frac{2\mu_s^2}{\lambda_m}\ ,\;\;\;\;
\lambda^2=\lambda_h\lambda_s-\frac{1}{4}\lambda_m^2=0\ .
\ee
If this happens, then the tree-level potential takes the simple form
\be
V=-\frac{1}{2}\mu_h^2\left(h^2+\frac{1}{2}\frac{\lambda_m}{\lambda_h}s^2\right)
+\frac{1}{4}\lambda_h 
\left(h^2+\frac{1}{2}\frac{\lambda_m}{\lambda_h}s^2\right)^2\ .
\ee
When thinking about further possible symmetries that could enforce such form of the potential one should keep in mind that this form is supposed to hold at some critical temperature, not at $T=0$. While the symmetric form of the quartic couplings would be approximately respected by thermal corrections (as quartics have only a logarithmic dependence on $T$), quadratic terms for $h$ and $s$ do evolve differently with temperature and would break that
symmetry. We will examine this in more detail in the next Subsection.

\begin{figure}[t]
\includegraphics[width=0.45\textwidth, clip ]{figs/flat_A.eps}
\hspace*{0.5cm}
\includegraphics[width=0.45\textwidth, clip ]{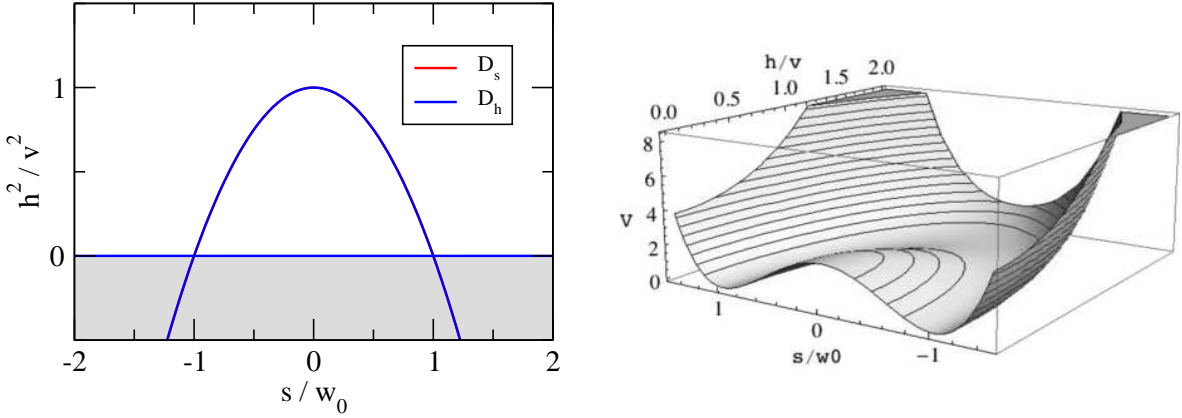}
\caption{\label{fig:flat1xx}
Special scenario with $\mathbf{Z}_2$ symmetry, $\lambda_m>0$ and $\lambda^2=0$ showing a 
flat direction. Left, degenerate parabolas $D_{h,s}^2(s)$. Right, 
corresponding potential.}
\end{figure}

There are two qualitatively-different cases depending on the sign of $\lambda_m$. (The limiting case $\lambda_m=0$ has little interest, as then $s$ and $h$ are completely decoupled from each other.)
For $\lambda_m>0$, the flat direction is the parabola
\be
h^2+\frac{1}{2}\frac{\lambda_m}{\lambda_h}s^2=\frac{\mu_h^2}{\lambda_h}\ ,
\ee
closed around the origin. Then the 
potential looks like a Mexican-hat potential, see Fig.~\ref{fig:flat1xx}. Of 
course such potential would not be acceptable at $T=0$ (implying in 
particular a massless scalar)
but could be of interest at $T=T_c$: the effect of the thermal cubic from 
gauge bosons can be enhanced by the flatness of the potential, leading to 
a large $v(T_c)/T_c$. This is confirmed by our numerical analysis. As we will 
see below, when $T$ falls below $T_c$ the minimum 
(which at $T=T_c$ is not located at any precise point along the flat 
direction) will be driven either to $h=0, s\neq 0$ (a case which does not interest us) or to $h\neq 0, s=0$, with a big jump in $v(T_c)/T_c$.

\begin{figure}[t]
\includegraphics[width=0.45\textwidth, clip ]{figs/flat_B.eps}
\hspace*{0.5cm}
\includegraphics[width=0.45\textwidth, clip ]{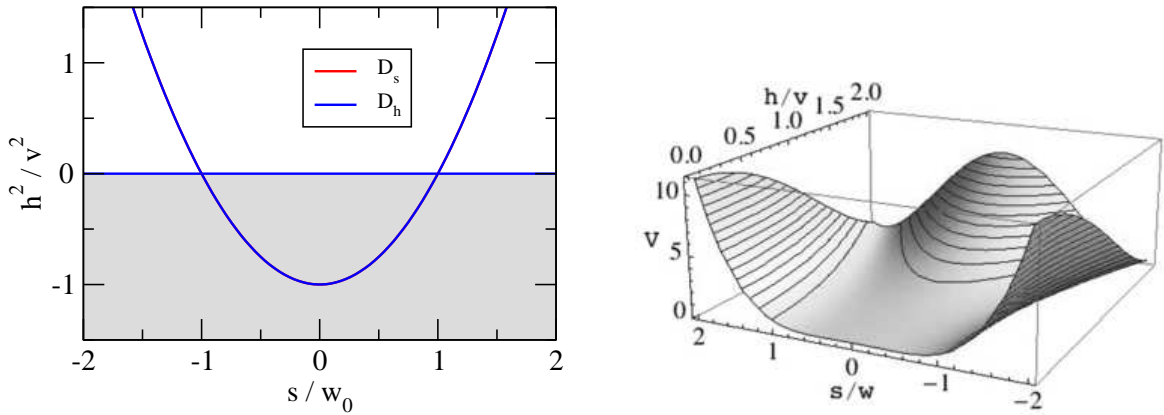}
\caption{\label{fig:flat2}
Special scenario with $\mathbf{Z}_2$ symmetry, $\lambda_m<0$ and $\lambda^2=0$ showing
two flat directions. Left, degenerate parabolas $D_{h,s}^2(s)$. Right, 
corresponding potential.}
\end{figure}

For $\lambda_m<0$, the degenerate parabolas correspond to two flat 
directions
\be
h^2-\frac{1}{2}\frac{|\lambda_m|}{\lambda_h}s^2= \frac{\mu_h^2}{\lambda_h}\ ,
\ee
running away to infinity, see Fig.~\ref{fig:flat2}. The 
stability of the potential along such directions should be ensured by 
mass terms or one-loop quartics.
This case could be of interest for the transition if the flat directions 
intersect $h=0$, which requires $\mu_h^2<0$.
The thermal evolution of the potential in this case will depend crucially on the 
thermal cubic.
\begin{figure}[t]
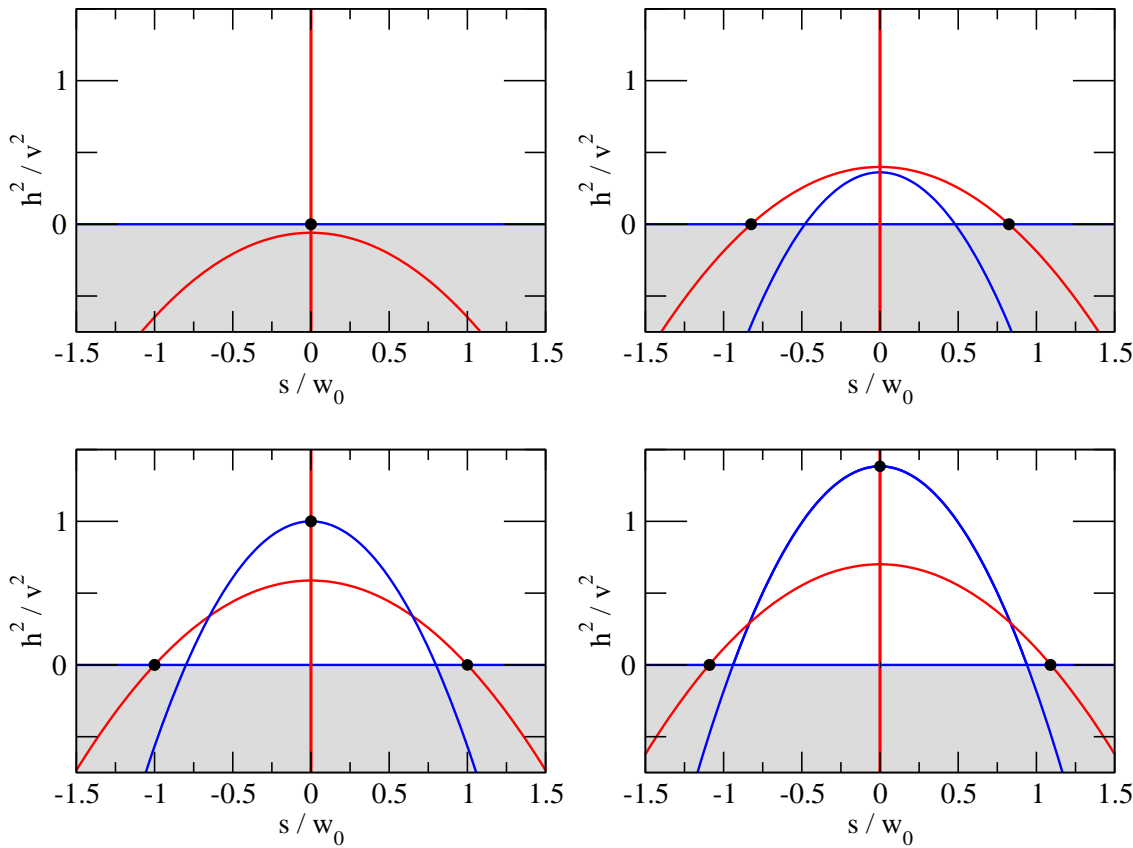

\begin{center}
\includegraphics[width=0.45\textwidth, clip]{figs/Z2T3.eps}
\includegraphics[width=0.45\textwidth, clip ]{figs/Z2T2.eps}
\end{center}
\begin{center}
\includegraphics[width=0.45\textwidth, clip ]{figs/Z2T0.eps}
\includegraphics[width=0.45\textwidth, clip ]{figs/Z2T1.eps}
\end{center}
\caption{\label{fig:Z2T}
Snapshots of the $T$-dependent curves $D_h^2(s)$ (blue lines) and $D_s^2(s)$ (red lines) intersecting at the $T$-dependent minima (black dots) in a $\mathbf{Z}_2$-symmetric scenario with the correct cosmological history. The plots are in order of decreasing $T$, from left to right and top to bottom, with $T\simgt T_Z$ (upper left); $T_Z> T>T_c$ (upper right); $T=T_c$ (lower left); and $T=T_c/2$ (lower right).}
\end{figure}

\subsection{Thermal Evolution and EW Phase Transition}

As we saw in Section~\ref{sec:strong_PT}, at leading order in the high-$T$
expansion, only $\mu_s^2$ and $\mu_h^2$ are affected, according to
eqs.~(\ref{massT}).\footnote{
Note that, in the context of the Singlet Majoron Model, right-handed neutrino Yukawas
give a significant contribution to $c_s$ and the imaginary component of the
(complex) singlet contributes to both $c_s$ and $c_h$, increasing the
coefficients of $\lambda_s$ and $\lambda_m$ in (\ref{thermalcs}).}
As a result, when $T$ is lowered from $T_c$ down
to $T=0$, the parabolas $D^2_h(s)$ and $D^2_s(s)$
simply drift away from $h=0$ keeping their width and symmetry axis
fixed. As they move at different rates, determined by [see eq.~(\ref{DT})] 
\be
\frac{d D_h^2(s)}{dT^2}=-\frac{c_h}{\lambda_h}\ ,\quad
\frac{d D_s^2(s)}{dT^)}=-\frac{2c_s}{\lambda_m}\ ,
\ee
their relative position can change, together with the location of the minima in the potential. This is illustrated by Fig.~\ref{fig:Z2T}, which shows snapshots of $D_{h,s}^2(s)$ at different temperatures. At very high $T$ the minimum is at $(0,0)$. The $\mathbf{Z}_2$ symmetry breaks spontaneously at some critical temperature $T_Z$ and the two minima at $(0,\pm w_0(T))$ move away 
from the origin as $T$ gets lower. Eventually the EW minimum forms
and gets degenerate with the $\mathbf{Z}_2$-breaking minima at $T_c$. For lower temperatures, the EW minimum is the deepest one. 

Let us focus on cases that lead to a tree-level barrier, which, as explained above,
require $w=0$, $\lambda_m>0$ and $ \lambda^2<0$. Once we identify the
parameters that give such barrier in the potential, $V_c(h,s)$, of the form (\ref{VcZ2}), we still have the freedom
to choose $T_c$ and to perform the appropriate rescaling to ensure $v(0)=v_{EW}$. 
To be specific, the potential is
\be
V = V_c(h,s) - \frac12 (T_c^2-T^2 ) (c_h h^2 + c_s s^2)\ .
\ee

As explained in Section~\ref{sec:strong_PT}, we start at $T_c$ with degenerate minima with broken and unbroken EW symmetry: $V_b(T_c)=V_s(T_c)$. As $T$ is lowered we want that the broken minimum gets deeper becoming our vacuum, in which
case, $w(T)$ will stay at zero for all $T<T_c$. One has
\be
\Delta V_{bs}(T)=V[v(T),0] - V[0,w_0(T)] = 
- \frac{\mu_h^4(T)}{4 \lambda_h} + \frac{\mu_s^4(T)}{4 \lambda_s}\ ,
\ee
so, to end up at the broken minimum at $T=0$  we need the condition\footnote{In this simple scenario, this condition coincides with the condition derived from $d\Delta V_{bs}(T)/dT^2=\left[c_h v^2-c_s w_0^2\right]/2>0$, see the discussion around eq.~(\ref{deepcond}).}
\be
\label{cond1}
\frac{c_h}{c_s}>\sqrt{\frac{\lambda_h}{\lambda_s}} = \frac{w_0^2}{v^2} \ ,
\ee
which can be taken as a constraint on the initial parameters $v$ and
$w_0$ in a specific model where $c_h$ and $c_s$ are known constants.
Alternatively, we can separate from $c_s$ its $\lambda_s$-dependent part [see (\ref{thermalcs})] as
\be
c_s=\frac{1}{4}\lambda_s+\delta c_s\ ,
\ee
where $\delta c_s=\lambda_m/6$ here, but in general can include contributions from other particles coupled to the singlet. Then, condition (\ref{cond1}) translates into a lower limit on $\lambda_s$:
\be
\label{lsmin}
\lambda_s>\lambda_{s,min}\equiv
\frac{4}{\lambda_h}\left[2c_h^2-\lambda_h\delta c_s-2c_h \sqrt{c_h^2-\lambda_h \delta c_s}\right]\ ,
\ee
while an upper limit follows from eq.~(\ref{lambdadegcondition}):
\begin{equation}
\lambda_s<\lambda_{s,max}\equiv \frac{\lambda_m^2}{4\lambda_h}\ .
\end{equation}
Obviously, $\lambda_{s,min}<\lambda_{s,max}$ should be satisfied. 

The $T$-dependence of our parametrization is as follows.
In the approximation we work, quartic couplings do not depend on the temperature while the rest of parameters do depend on it, leading to $m_h^2(T)$, $m_s^2(T)$ and $v(T)$.
This $T$-dependence can be extracted from eqs.~(\ref{reli})-(\ref{relf}), after feeding in them the $T$-dependent $\mu_h^2$ and $\mu_s^2$. In this way it is straightforward to extract
\be
\label{v2T}
v^2(T)= v_{EW}^2-\frac{c_h}{\lambda_h}T^2\ ,
\ee
where we are always implicitly assuming $T<T_c$ (so that the EW minimum is the global one) which is the range of interest to run parameters from $T_c$ down to $T=0$.
From this we can already extract the important ratio
$v(T_c)/T_c$ as
\be
\frac{v(T_c)}{T_c}=\sqrt{\frac{v^2_{EW}}{T_c^2}-\frac{c_h}{\lambda_h}} \ .
\label{voverT}
\ee
Notice, however, that the phase transition cannot be made arbitrarily strong by choosing
low $T_c$, in which case $v(T_c)\simeq v_{EW}$, since then the tunneling probability becomes small and
the meta-stable symmetric phase will become stable. Moreover, for low $T_c$
the high-$T$ approximation breaks down. 

Eq.~(\ref{voverT}) becomes more meaningful when combined with information
on the mass spectrum.  For the Higgs mass parameter we have
\be
m_h^2(T)=2\lambda_h v^2(T)\ ,
\ee
where the physical value of the Higgs mass $M_h^2 = m_h^2(0)$ fixes $\lambda_h$ through $M_h^2=2\lambda_h v_{EW}^2$ [here we use capital letters for the $T=0$ parameters: $M_h^2\equiv m_h^2(0)$ and $M_s^2\equiv m_s^2(0)$]. Similarly, the singlet mass is given by
\be
m_s^2(T)=m_s^2+\left(\frac{\lambda_m}{2\lambda_h}c_h-c_s\right)(T_c^2 - T^2)\ .
\ee
Notice that the second term is positive for $T<T_c$ due to
$\lambda_h\lambda_s<\lambda_m^2/4$ and eq.~(\ref{cond1}), such that the zero temperature
mass of the singlet is larger than the one at the critical temperature:
$M_s^2 > m_s^2$. From the condition of degenerate minima,
eq.~(\ref{degconds}), we can obtain the additional relation
\be
m_s^2=\frac{1}{2}v^2(T_c)(\lambda_m-2\sqrt{\lambda_h\lambda_s})\ ,
\ee
arriving at
\be
\label{singletmass}
M_s^2=\frac{1}{2}\left(\lambda_m-2\sqrt{\lambda_h\lambda_s}\right)v_{EW}^2
+\left(c_h\sqrt{\frac{\lambda_s}{\lambda_h}}-c_s\right)T_c^2\ .
\ee
So, the singlet mass squared is a simple linear combination of the two
mass scales $v_{EW}^2$ and $T_c^2$ with positive coefficients. For
$\lambda_s=\lambda_{s,min}$, the coefficient of $T_c^2$ in
eq.~(\ref{singletmass}) is zero and $M_s$ is independent
of $T_c$. In the case $\lambda_s=\lambda_m^2/(4\lambda_h)$ it is the
coefficient of $v_{EW}^2$ that cancels, and then $M_s$ increases
linearly with $T_c$. This is precisely the limiting case with a flat direction
at $T_c$ (of $\lambda_m>0$ type) discussed in the previous Subsection.

\begin{figure}[t]
\begin{center}
\includegraphics[width=0.9\textwidth, clip ]{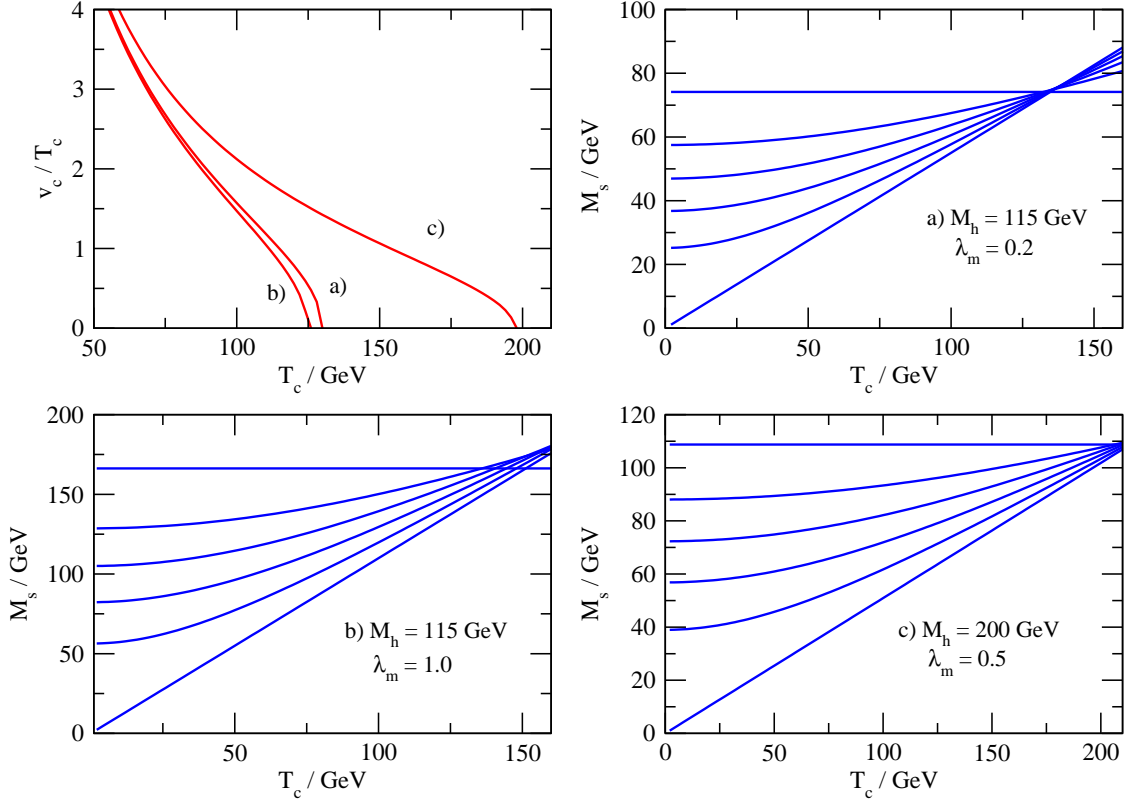}
\end{center}
\caption{\label{fig:Z2results}
Ratio $v(T_c)/T_c$ (upper-left plot) and values of the 
singlet scalar mass (rest of plots) as a function of the critical 
temperature for the cases $a)$ $M_h=115$ GeV, $\lambda_m=0.2$;
$b)$ $M_h=115$ GeV, 
$\lambda_m=1$; $c)$ $M_h=200$ GeV and $\lambda_m=0.5$. Different masses 
correspond to different values of 
$\lambda_s\in(\lambda_{s,min},\lambda_{s,max})$, with $M_s$ increasing for lower $\lambda_s$. 
}
\end{figure}

To explore in more detail what masses and $v(T_c)/T_c$ are allowed we will proceed as follows. Cases with a
barrier are simply found by choosing $\lambda_m>0$ and $-\lambda^2$ in
the interval $(0,\lambda_m^2/4)$, which in terms of
$\lambda_s$ is equivalent to $\lambda_s\in(0,\lambda_{s,max})$. The lower
part of this interval is removed by the condition (\ref{lsmin}),
required to guarantee the correct $T\rightarrow 0$ limit, leaving only
the interval $(\lambda_{s,min},\lambda_{s,max})$.  Now, for fixed $M_h$ and using $T_c$ as a
parameter, we can obtain $v(T_c)/T_c$ and $M_s$ for different choices
of $\lambda_m$ and $\lambda_s$ in the appropriate ranges described
above.  The results are as shown in Fig.~\ref{fig:Z2results}, which presents three representative cases: $a)$ $M_h=115$~GeV with $\lambda_m=0.2$ and then $\lambda_s\in(0.03,0.09)$; $b)$ $M_h=115$~GeV with $\lambda_m=1$ and $\lambda_s\in(0.09,2.29)$;
and $c)$ $M_h=200$~GeV with $\lambda_m=0.5$ and $\lambda_s\in(0.09,0.19)$. In the upper plot, we show the
large value of $v(T_c)/T_c$ (which is independent of $\lambda_s$) that can be obtained as a function of $T_c$. For $v(T_c)/T_c\simgt 4$ we cannot trust the high-$T$ approximation, so we do not explore smaller values of $T_c$. In the rest of the plots we show the value of the singlet scalar mass $M_s$ as a function of the
critical temperature $T_c$ for equally spaced values of
$\lambda_s\in(\lambda_{s,min},\lambda_{s,max})$. Higher
singlet masses correspond to lower values of $\lambda_s$. One concludes
that very strong
first-order EW transitions can be obtained for a wide range of scalar masses.

\subsection{One-loop Numerical Analysis}

\begin{figure}[t]
\begin{center}
\includegraphics[width=0.7\textwidth, clip ]{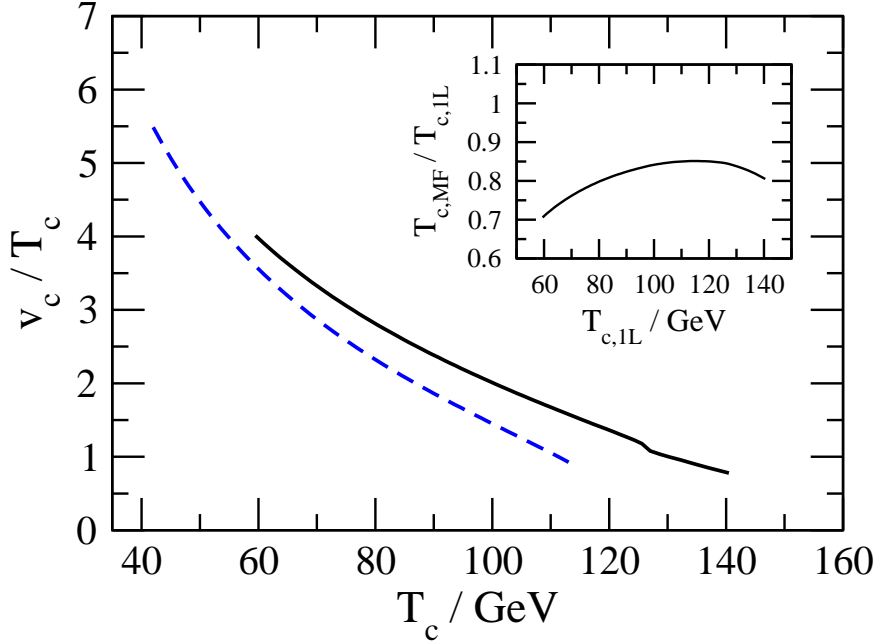}
\end{center}
\caption{\label{fig:Z2oneloop}
The ratio $v_c/T_c$ in the mean-field approximation (blue dashed line) and the one-loop approximation (black solid line) as a function of the corresponding critical $T_c$. The relation between critical temperatures in the two approximations is shown in the inset. Here $M_h=115$ GeV, $M_s=145$ GeV and $w_0=100$ GeV.  $T_{c,MF}$ corresponds to the critical temperature in the mean-field approximation, while $T_{c,1L}$ takes into account the full one-loop resummed potential.}
\end{figure}

So far, we have identified choices for the $T=0$ parameters that lead to strong electroweak phase transitions in the mean-field approximation. It is straightforward to refine these results starting from the same tree-level parameters but including in the scalar potential one-loop $T=0$ corrections and the full one-loop thermal integrals (which correctly take into account Boltzmann decoupling effects) further improved by daisy resummation. Details of this standard procedure are given in Appendix~A. To illustrate the impact of this refinement, we show in Fig.~\ref{fig:Z2oneloop} the ratio $v_c/T_c$ in the mean-field approximation (blue dashed line) compared with the same quantity calculated with the one-loop thermal potential just described (black solid line). The example shown corresponds to $M_h=115$ GeV and $M_s=145$ GeV. As expected from
the correct inclusion of Boltzmann decoupling effects, which tend to increase $T_c$ (see inset, where $T_{c,MF}$ corresponds to the critical temperature in the mean-field approximation, while $T_{c,1L}$ takes into account the full one-loop resummed potential), the one-loop transition is weaker\footnote{When comparing the two curves in Fig.~\ref{fig:Z2oneloop}, keep in mind that, for a given choice of the model parameters, the resulting $v(T_c)/T_c$ at one-loop is displaced to higher $T_c$ and lower $v(T_c)/T_c$ with respect to the tree-level value (like the curves' end-points demonstrate).} than in the mean-field approximation but still strong enough to allow for baryogenesis. 

We have also examined at one-loop cases with a flat direction with $\lambda_m>0$, case in which thermal cubics from bosons play an important role. Although such thermal effects would not give rise to strong transitions by themselves (for weak couplings to the Higgs field), in the presence of a flat direction the thermal cubic lifts the flat direction creating a barrier between  the broken and symmetric minima, and  ensures large jump in $v(T)/T$, leading to strong phase transitions at one-loop. Concerning the naturalness of such scenario,
notice that all that is required to realize it is that quartic couplings satisfy $\lambda^2=0$ (which might be the result of some symmetry). This makes 
the parabolas $D_h^2(s)$ and $D_s^2(s)$ equally wide and, as they are both centered 
at $s=0$, when they shift with temperature, it is guaranteed that they will overlap and
give a flat direction at some $T_c$.

Our results contradict some claims in the literature concerning this scenario \cite{Barger:2007im}, which were focused on transitions driven by the thermal cubic.  On the other hand, the analysis of the Singlet Majoron model in \cite{EWPHTMajoron} did find strong transitions. One important difference between that study and ours is that, for phenomenological reasons (in particular to be able to generate nonzero neutrino masses),
the scenarios considered in \cite{EWPHTMajoron}\footnote{A direct comparison of results is difficult because the analysis in \cite{EWPHTMajoron} is based on  scans of the parameter space (for instance, from distributions of $\lambda_h, \lambda_s$ and $\lambda_m$ we cannot obtain the distribution of $\lambda^2$, which is a crucial parameter for EWPhTs) and different mechanisms operate in different regions of parameter space.} are restricted to $w\neq 0$, which works against the possibility of obtaining a really strong phase transition through tree-level barriers. 
However, loop effects related to sizable Yukawa couplings (to the right-handed neutrinos), that we are not discussing, can help in getting strong EWPhTs. 
On the other hand, it is clear that the example in Fig.~9 of \cite{EWPHTMajoron} corresponds to a one-loop deformation of a tree-level case with a nearly flat direction. One-loop effects
shift the EW minimum away from $w=0$ and lead to a very light scalar, see eq.~\ref{Z2blight}.

\section{Special Cases: A Supersymmetric Example\label{sec:nmssm}}
A different context in which a particular region of parameter space is selected by some mechanism, is that of supersymmetry.
A supersymmetric version of the SM plus singlet model, the Next to Minimal Supersymmetric Standard Model (NMSSM), has been considered since long ago as suitable to obtain a strong electroweak phase transition with applications to electroweak baryogenesis.
In particular, ref.~\cite{Pietroni} was the first to emphasize in this context the relevance of a tree-level cubic term in the scalar potential.

Here we will examine a particular deformation of this model, the near-to-Minimal Supersymmetric Standard Model (nMSSM) \cite{nMSSM}, which differs from the NMSSM
in having a singlet superpotential with a loop-suppressed tadpole and no cubic term.
Refs.~\cite{EWPHTnMSSM} studied the electroweak phase transition in this model finding strongly
first-order cases. In the region of parameter space examined in \cite{EWPHTnMSSM} the scalar potential reduces to that of a SM Higgs plus a real scalar with a potential of our general form (\ref{Vtree})  but with
\be\label{con000}
\lambda_s=0\ , \quad \mu_3=0\ ,\quad \lambda_m>0\ .
\ee
It is interesting to note that a shift of the singlet respects the conditions $\mu_3=0$ and $\lambda_s=0$ so that there
is freedom to set $\mu_m=0$ without loss of generality. In terms of the parameters of eqs.~(\ref{reli})-(\ref{relf}) these conditions read
\be
\lambda^2=-\frac{\lambda_m^2}{4},\quad m_*=-\frac{\lambda_m^2}{4}w,\quad m_{sh}^2=\lambda_m v w
\ee
and the model can be described by just 5 parameters, 
\be
\label{eq:params2}
\{v,w, m_{h}^2, m_{s}^2, \lambda_m \}\ .
\ee

\begin{figure}[t]
\includegraphics[width=0.45\textwidth, clip ]{figs/Ds_nmssm.eps}
\hspace*{0.1cm}
\includegraphics[width=0.55\textwidth, clip ]{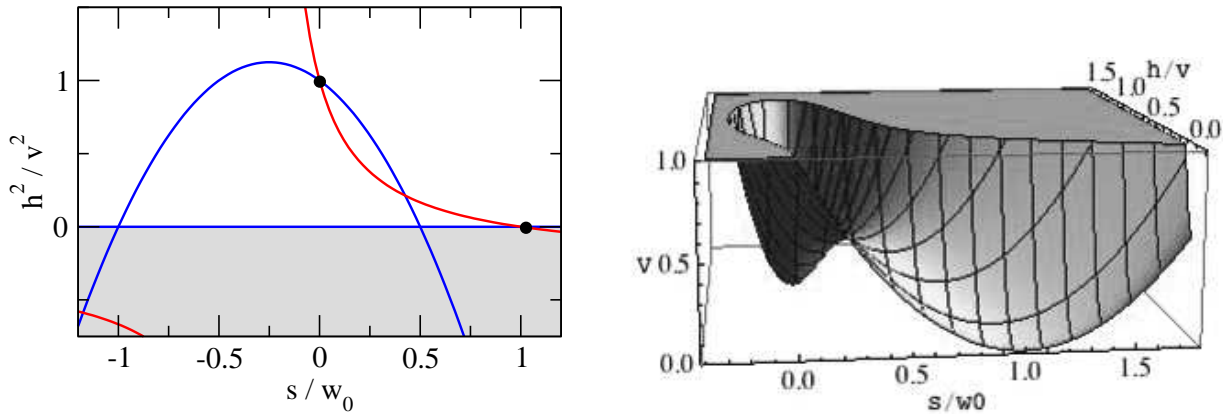}
\caption{\label{fig:nMSSM}
Scenario with a strong transition in the nMSSM model. Left, intersecting curves $D_{h,s}^2(s)$. Right, 
corresponding potential.}
\end{figure}

The model has sufficient structure in the $D_{h,s}^2(s)$ functions to allow for tree-level
barriers, which must necessarily be of type-(d)  in the classification of eqs.~(\ref{casea})-(\ref{cased}), that is, of those with $\lambda_m>0$ (imposed by supersymmetry) and $(w_0-w_p)(w-w_p)>0$. This last condition can be explicitly checked after translating the conditions of eq.~(\ref{con000}) into the parameters for a degenerate minimum (see Table~\ref{Table1}) or, equivalently, imposing the parameter constraints needed to have a symmetric minimum at $(0,w_0)$ degenerate with the broken one at $(v,w)$,
which now read:
\be
w_0=w+\frac{m_h^2 v^2}{8 m_s^2 w}\left[1+\sqrt{1+\left(4\frac{m_sw}{m_hv}\right)^2}\right]\ ,\quad
\lambda_m=\frac{m_h^2}{w^2}\left[-1+\sqrt{1+\left(\frac{4m_s w}{m_h v}\right)^2}\right]\ ,
\ee
and $w_p=0$, from which one gets $(w_0-w_p)(w-w_p)=ww_0$, which is always positive. Fig.~\ref{fig:nMSSM} shows an example of such barrier.

The potential with degenerate vacua takes the simple form
\be
V = \frac18 v^2 m_h^2 \left[
\left( \frac{h^2}{v^2} +\frac{s-w}{w_0}-1 \right)^2
+ \left(4 \frac{R_s}{R_h} -1 \right)
\frac{h^2 (s-w)^2}{v^2 w_0^2}\right]\ ,\label{VdnMSSM}
\ee
where we have one single shape parameter: $R_s/R_h$, as in the $\mathbf{Z}_2$-symmetric scenario, but now with $\lambda_m/R_h=2R_s/R_h-1/2$ and $R_{sh}/R_h=1/2$.
In the particular case $4R_s/R_h=1$ the potential (\ref{VdnMSSM})
develops a flat direction which, in the  $(h^2/v^2,s/w_0)$-plane,  is a straight line through both minima. In terms of the original quartic couplings of the potential this scenario corresponds 
to the extreme limit $\lambda_m=0$, that we do not consider further.

\subsection{Thermal Evolution and EW Phase Transition}

As in previous scenarios, we start from the potential with degenerate minima, which is assumed to hold at some $T_c$, and then use the mean-field approximation to derive the corresponding potential parameters at $T=0$. This temperature evolution 
is given in eq.~(\ref{massT}) where now the constants $c_h$, $c_s$ and $m_3$ will also receive contributions from supersymmetric particles (if their masses are not much higher than $T$). As the
model can support a strong electroweak phase transition without
the need of supersymmetric particles coupled sizeably to the Higgs (as needed in the MSSM case) we simplify the analysis by assuming that thermal effects from superpartners are Boltzmann suppressed.
If supersymmetric particles do not decouple from the thermal plasma their effects can be included and they will only modify quantitatively our discussion of strong phase transitions based on
the tree-level potential.

Note that the condition $\mu_m=0$, that we have previously imposed using the shift symmetry, is respected by thermal corrections
(in the mean-field approximation) and implies that $D_h^2(s)$
is centered at $w_p=0$ and has the same width at all $T$'s while its vertex $\bar h$ will evolve with $T$ as in the $\mathbf{Z}_2$ case. Another good property of this choice is that $m_3=0$, see (\ref{thermalcs}), so that $\mu_1^3$ is also $T$-independent. This makes the evolution of $D_s^2(s)$ also very simple:
\be
\frac{D^2_s(s)}{dT^2}=-\frac{2c_s}{\lambda_m}\ ,
\ee
exactly as in the $\mathbf{Z}_2$ case, so that also $D_s^2(s)$ keeps its shape and simply drifts. As usual, to have a successful cosmological history, $D_{h,s}^2(s)$ should move in such a way that the EW vacuum forms and gets deeper at low temperature. 
\begin{figure}[t]
\begin{center}
\includegraphics[width=0.7\textwidth, clip ]{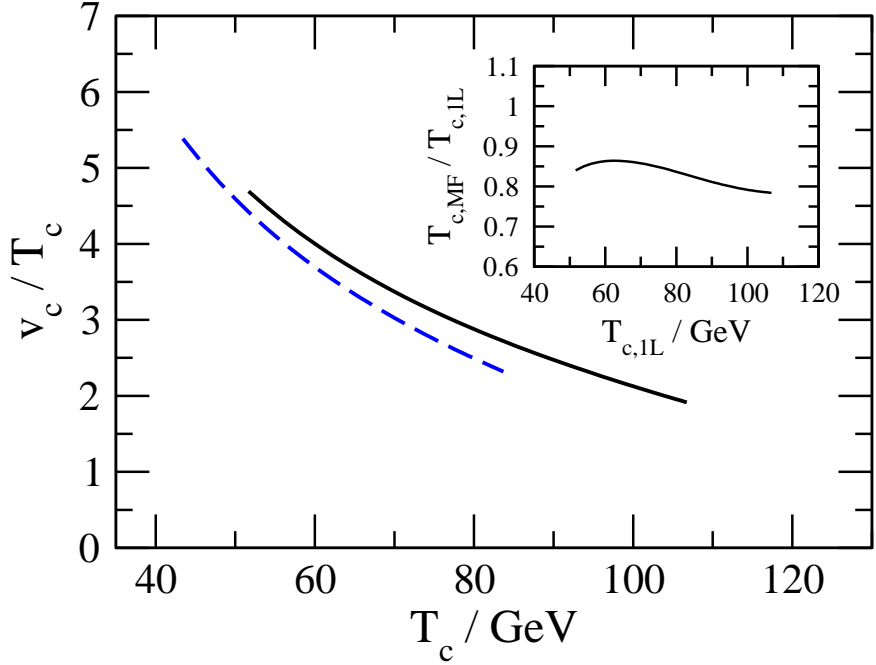}
\end{center}
\caption{\label{fig:nMSSMvoT}
The ratio $v_c/T_c$ in a nMSSM scenario in the one-loop approximation (black solid line) as a function of the corresponding critical $T_c$. The relation between critical temperatures in the two approximations is shown in the inset. Here $M_h=135$ GeV, $M_s=145$ GeV and $w_0=-400$ GeV.}
\end{figure}

The thermal evolution of the potential minimum in this model is more complicated than in the $\mathbf{Z}_2$ symmetric case:
$v(T)$, $w(T)$ and $m_s(T)$ are determined by the simple equations:
\bea
\mu_h^2(0) - c_h T^2 & = & \lambda_h v^2(T)+\frac{1}{2}\lambda_m w^2(T)\ , \\
 \mu_s^2(0) + c_s T^2 & = & m_s^2(T) - \frac{1}{2}\lambda_m v^2(T)\ ,\\
\mu_1^3 & = & - m_s^2(T) w(T) = \mu_s^2(T)w_0(T)\ ,
\eea
while $m_h^2(T)$ is trivially related to $v(T)$ by $m_h^2(T)=2\lambda_h v^2(T)$, as usual.
This system of equations can be solved analytically, although the resulting expressions for $v^2(T)$, $w(T)$ and $m_s^2(T)$, being solutions of cubic equations, are not very illuminating and we refrain from writing them down.

Once the temperature dependence of all parameters is known, we can  relate the parameters at the critical temperature
\be
\{v, w, m_{h}^2, m_{s}^2, \lambda_m \}
\ee
to those at zero temperature 
\be
\{v_{EW}, w_{EW}, M_{h}^2, M_{s}^2, \lambda_m \}
\ee
and, in particular, determine the critical temperature in terms of the physical parameters. 
Some simple relations are
\bea
M_h^2  & = & m_h^2 \frac{v^2_{EW}}{v^2}\ , \nonumber\\
M_s^2 & = & m_s^2 \frac{w}{w_{EW}}\ ,\\
M_s^2 - m_s^2 & = &\frac12 \lambda_m (v_{EW}^2 - v^2)-
c_s T_c^2 \ .\nonumber
\eea
Once we have the $T=0$ potential we can perform a refined one-loop analysis including thermal decoupling effects as already discussed in previous Sections to confirm the existence of strong phase transitions in the regions indicated by the tree-level analysis.
An example of the results we obtain is given in Fig.~\ref{fig:nMSSMvoT}.

\section{Special Cases: Light Scalar\label{sec:dark}}

As a final  example we consider a realization of the SM plus a singlet  with a very light scalar,  put forward in \cite{Fox} as a possible way of increasing the strength of the EWPhT. The model has
\be
\lambda_m=0\ ,\quad \mu_1=0\ , \quad \mu_3=0\ , \quad \lambda_s=0\ ,
\ee
and $\mu_m$, the only coupling connecting the $s$ and $h$ sectors, is small: $\mu_m\equiv\epsilon_m v_{EW}\ll v_{EW}$. The condition $\mu_1=0$ is not respected by thermal corrections and therefore we have to keep it nonzero in our discussion, but with $\mu_1(0)=0$. In our parametrization this reads 
\be
\lambda_m=0\ ,\quad \lambda^2=0\ , \quad m_*=0.
\ee
This case has only five free parameters that we can take as $\{v,w,m_h^2,m_s^2,\mu_m\}$. So, this is a very constrained scenario  and one sees (cf. Table~\ref{Table1}) that the conditions (\ref{condDET})-(\ref{CondH}) cannot be fulfilled: it is not possible to have degenerate minima with a barrier in between. This can be easily understood by studying the functions $D_{h,s}^2(s)$, which are  straight lines now:
\bea
D_h^2(s)&=&\frac{1}{2\lambda_h}(2\mu_h^2-\mu_m s)\ ,\\
D_s^2(s)&=&-4\frac{\mu_1^3+\mu_s^2s}{\mu_m}\ ,
\eea
and, with such simple structure, cannot lead to a tree-level barrier unless
a scenario with a flat direction (or close to it) is realized.
Such flat direction requires $\mu_m\mu_h^2=-4\lambda_h\mu_1^3$ and $\mu_m^2=8\lambda_h\mu_s^2$.
To decide how natural this is, we need to examine the thermal evolution of these quantities.

\begin{figure}[t]
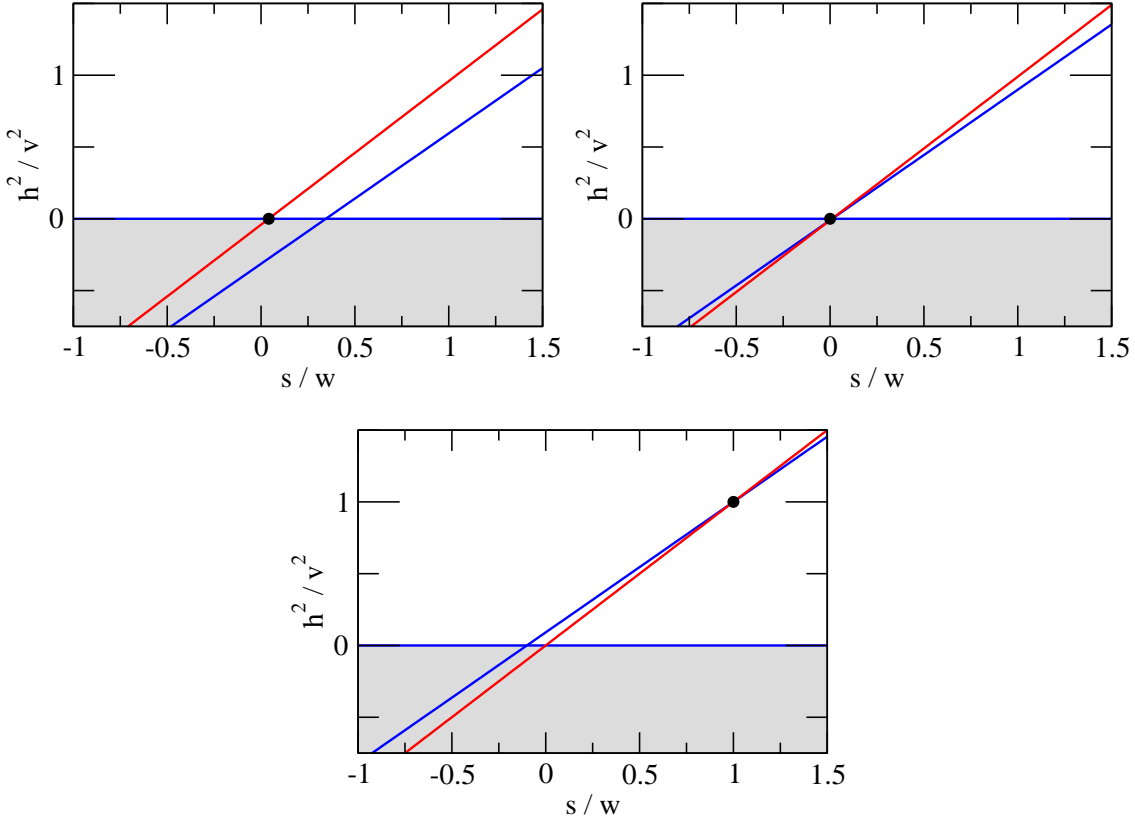

\begin{center}
\includegraphics[width=0.45\textwidth, clip ]{figs/LightT2.eps}
\includegraphics[width=0.45\textwidth, clip ]{figs/LightT1.eps}
\end{center}
\begin{center}
\includegraphics[width=0.45\textwidth, clip ]{figs/LightT3.eps}
\end{center}
\caption{\label{fig:LightT}
Snapshots of the $T$-dependent lines $D_h^2(s)$ (blue) and $D_s^2(s)$ (red) intersecting at the $T$-dependent minimum (black dot) in a scenario with a light singlet having the correct cosmological history. The plots are in order of decreasing $T$, from left to right and top to bottom, with $T> T_c$ (upper left); $T=T_c$ (upper right); and $T<T_c$ (bottom).}
\end{figure}
\subsection{Thermal Evolution and EW Phase Transition}

\begin{figure}[t]
\begin{center}
\includegraphics[width=0.7\textwidth, clip ]{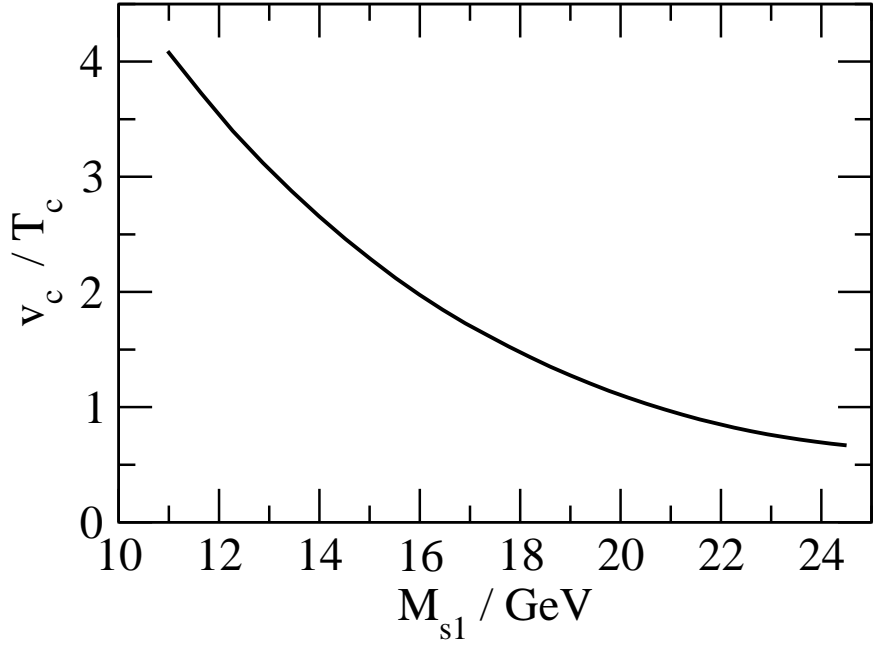}
\end{center}
\caption{\label{fig:dark}
The ratio $v_c/T_c$ in the scenario with a light scalar of Section~\ref{sec:dark}  in the  one-loop approximation as a function of the light scalar mass. Here $M_h\simeq M_{s2}\simeq 127$ GeV and $\mu_m=-40$ GeV.}
\end{figure}

In this model,  $\mu_s^2$ is $T$-independent in the mean-field approximation ($c_s=0$),
while $\mu_h^2$ and $\mu_1^3$ depend on $T$ in the usual way, now with $m_3=\mu_m/12$. The evolution of the two curves $D_{h,s}^2(s)$ with temperature is then quite simple: 
\be
\frac{dD_h^2(s)}{dT^2}=-\frac{c_h}{\lambda_h}\ ,\quad
\frac{dD_s^2(s)}{dT^2}=-\frac{1}{3}\ .
\ee
As seen in the $(h^2,s)$-plane, when $T$ increases, $D_{h,s}^2(s)$ shift towards the axis $h=0$ without rotating.
In order to have a proper cosmological evolution, starting in the symmetric vacuum at high $T$ and ending up at $T=0$ in the broken
EW minimum, the line $D_h^2(s)$ should have slope smaller than that of $D_s^2(s)$
(this requires $8\lambda_h\mu_s^2>\mu_m^2$) and it should move faster with $T^2$
than $D_s^2(s)$ (this requires $c_h>\lambda_h/3$, which is automatically satisfied).  An example to illustrate this scenario is presented in Fig.~\ref{fig:LightT}. In the 
mean-field approximation the EWPhT would be second-order. However, a large $v(T_c)/T_c$ could be achieved after including in the potential the full one-loop corrections (that include a cubic term for $h$ not only through transverse gauge bosons but also through the Higgs itself) if the model is tuned to have a nearly flat direction at $T_c$. The tuning involved, which we quantify by the small parameter $0<\epsilon\ll 1$,  requires nearly equal slopes for $D_{h,s}^2(s)$:
\be
8\lambda_h \mu_s^2 = \mu_m^2 (1+ \epsilon) =\epsilon_m^2(1+\epsilon)v_{EW}^2\ .
\label{tuning}
\ee
In terms of more physical parameters this reads
\be
4M_h^2M_s^2\simeq \epsilon_m^2 v_{EW}^4\ ,
\ee
which indeed requires a light scalar singlet, corresponding to the field excitations along the flat direction.  Indeed, including the effects of $s$-$h$ mixing, the two scalar mass eigenvalues are 
\be 
M_{s1}^2\simeq M_h^2\ ,\quad M_{s2}^2\simeq \frac{1}{4}\epsilon_m^2 \epsilon\frac{v_{EW}^4}{M_h^2}\ .
\ee 
Although it is difficult to imagine a symmetry reason that could lead to the relation (\ref{tuning}), once that tuning is arranged, temperature corrections do not spoil it.
In the mean-field approximation the critical temperature is easily computed to be
\be
T_c^2=\frac{\lambda_h\epsilon v_{EW}^2}{c_h(1+ \epsilon) -\lambda_h/3}\ ,
\ee
 which shows that $T_c$ can be made much smaller than in the SM, helping to increase  $v(T_c)/T_c$. To determine this quantity one needs to carry out the one-loop analysis. Our renormalization conditions for this particular scenario are detailed in Appendix~\ref{app:thermalpot}. Fig.~\ref{fig:dark}
shows our results. We have varied $\epsilon$ between $0.08$ and $0.2$, which roughly corresponds to light scalars with masses $M_{s_1}=10\div 25$ GeV, and we have fixed $\mu_m=-40$ GeV (this parameter has little influence on $v_c/T_c$) and
$M_h=120$ GeV (leading to $M_{s_2}\simeq 127$ GeV). We see that quite strong EWPhTs are possible, with $v_c/T_c$ increasing with decreasing $\epsilon$ (or $M_{s_1}$), which makes the flat-direction flatter.

\section{Numerical Examples. General Case\label{sec:general}}  

From the general analysis of the tree-level potential in Section~\ref{sec:PhT} we have learned that the SM with a singlet has a very rich structure. In particular we showed in eqs.~(\ref{casea})-(\ref{cased}) that there are four distinct types of arranging for two degenerate minima in the potential, one of which, at $(v,w)$, breaks the electroweak symmetry while the other, at $(0,w_0)$, does not. The potential with these degenerate vacua is given by eq.~(\ref{Vdgen}).

\subsection{Thermal Evolution and EW Phase Transition}

As already explained in the particular cases studied in previous Sections, we take the previous potential with degenerate minima to hold at some critical temperature $T_c$ and use the mean-field
approximation to the free-energy to obtain the corresponding
tree-level potential at $T=0$. The temperature dependence of
the potential parameters is very simple but finding how this
dependence affects the minima, in particular $v(T)$ and 
$w(T)$, requires solving a cubic equation. A smart choice of the 
singlet shift can simplify this task by leading to simpler 
analytical expressions for these quantities. Here we simply 
perform this thermal evolution of parameters down to $T=0$ numerically.

We start from one particular potential with degenerate minima, like those just discussed, expressed in terms of a set of original parameters $\{\mu_h^2, \mu_s^2,\mu_m, \mu_3, \mu_1^3,\lambda_h, \lambda_s, \lambda_m\}$. If we rescale all mass parameters by the appropriate 
power of some factor $A(T_c)$ we still have one potential with degenerate minima corresponding to the set
\be 
\{A(T_c)^2\mu_h^2,A(T_c)^2 \mu_s^2,A(T_c)\mu_m,A(T_c)\mu_3, A(T_c)^3\mu_1^3,\lambda_h, \lambda_s, \lambda_m\}\ .
\ee
The thermally corrected potential (in mean-field approximation) for that rescaled potential reads
\bea
V_T(h,s) &=& -\frac{1}{2}A(T_c)^2\mu_h^2 h^2+\frac{1}{4}\lambda_h h^4 +\frac{1}{2}A(T_c)^2\mu_s^2 s^2 +\frac{1}{4}\lambda_s s^4+
\frac{1}{4}A(T_c)\mu_m s h^2+\frac{1}{4}\lambda_m s^2h^2\nonumber \\
&&+A(T_c)^3\mu_1^3 s+\frac{1}{3}A(T_c)\mu_3 s^3+\left[\frac{1}{2}c_h h^2 +\frac{1}{2}c_s s^2+A(T_c)m_3 s\right](T^2-T_c^2)\ ,
\label{VTgen}
\eea
with $c_h, c_s$ and $m_3$ as given in eq.~(\ref{thermalcs}). The factor $A(T_c)$ affects all dimensionful parameters and is used to guarantee $v(0)=v_{EW}$. The function $A(T_c)$ and the singlet vacuum expectation value at $T=0$, $w_{EW}$, are obtained by solving numerically the minimization equations $\partial V_T/\partial h=0$ and $\partial V_T/\partial s=0$ at $T=0$, $h=v_{EW}$, $s=w_{EW}$. Once $A(T_c)$ is known as a function of $T_c$, eq.~(\ref{VTgen}) describes at $T=0$ a family of potentials
(parametrized by $T_c$) that lead to a strong electroweak phase transition.

\begin{figure}[t]
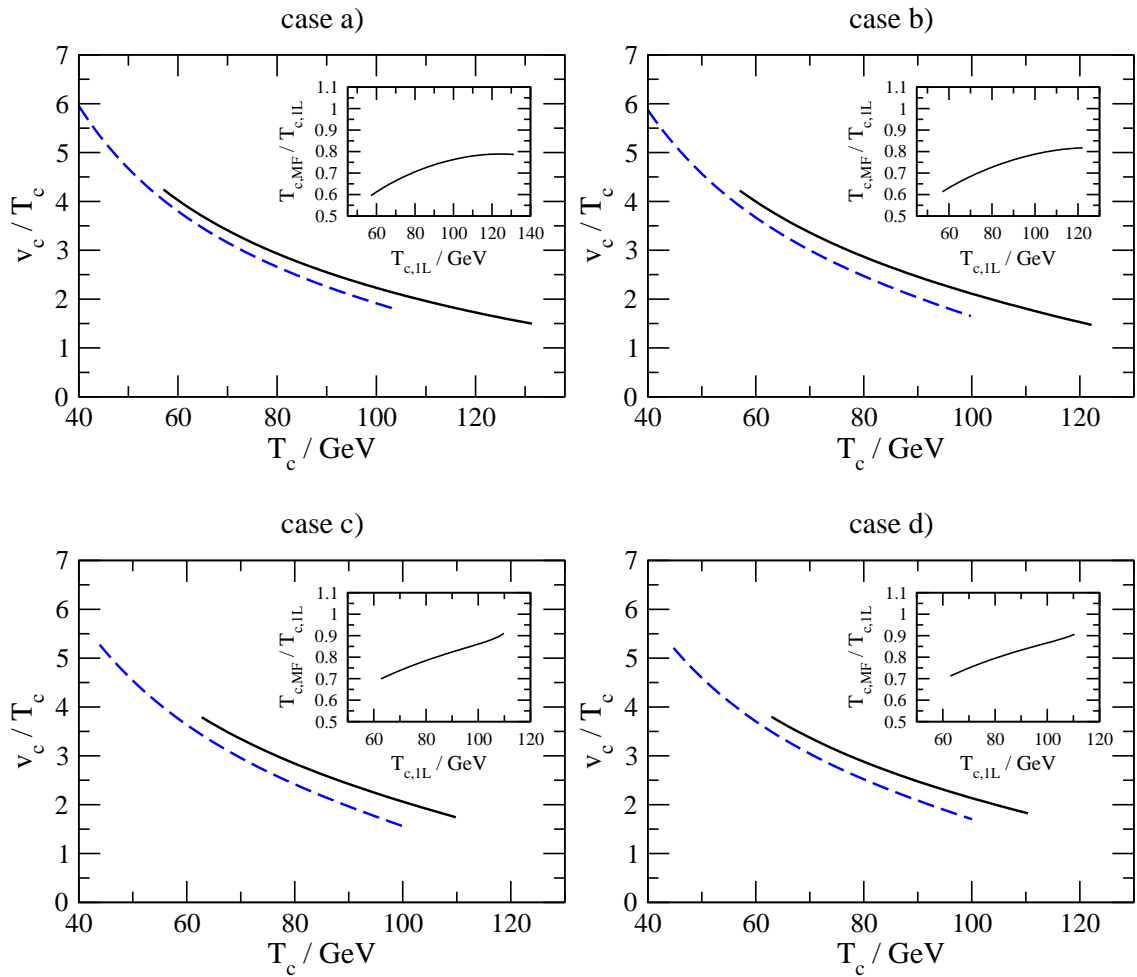

\begin{center}
\includegraphics[width=0.45\textwidth, clip ]{figs/voT_caseA.eps}
\includegraphics[width=0.45\textwidth, clip ]{figs/voT_caseB.eps}
\end{center}
\begin{center}
\includegraphics[width=0.45\textwidth, clip ]{figs/voT_caseC.eps}
\includegraphics[width=0.45\textwidth, clip ]{figs/voT_caseD.eps}
\end{center}
\caption{\label{fig:voTgen}
The ratio $v_c/T_c$  in the mean-field approximation (blue dashed line) and the one-loop approximation (black solid line) as a function of the corresponding critical $T_c$ for generic scenarios of different types, as indicated. The relation between critical temperatures in the two approximations is shown in the insets. }
\end{figure}

Such potentials can be taken as the starting point for a full one-loop analysis of the EWPhT, including one-loop corrections at $T=0$ and finite temperature, see Appendix~\ref{app:thermalpot}. We have performed such analysis for some representative families of potentials of the four different types discussed in Section~\ref{sec:PhT}. For our numerical examples we have chosen the
parameters as given in Table~\ref{tab:gen_ex}, where $\alpha_{sh}$ is the scalar mixing angle. The results are presented in Fig.~\ref{fig:voTgen}, which shows the ratio $v(T_c)/T_c$ as a function of $T_c$ both in mean-field approximation (dashed line) and one-loop (solid line). The critical temperatures in both approximations differ and we also plot their ratio $T_{c,MF}/T_{c,1L}$ in the insets of each plot. Although the generic effect of including the full one-loop thermal corrections is always to increase the critical temperature and lower $v(T_c)/T_c$, Fig.~\ref{fig:voTgen} shows that very strong transitions can be obtained, with $v(T_c)/T_c>1$ as required for
successful baryogenesis, in all cases.

\begin{table}
\begin{center}
\begin{tabular}{| c  || c | c | c | c | c | c |}
\hline
case & $M_{s1}$ / GeV & $M_{s2}$ / GeV & $\sin(\alpha_{hs})$ &
$\lambda_m$ & $w$ / GeV & $w_0$ / GeV \\
\hline \hline
 a) & $115.5$ &  $209.3$ & $0.612$ & $-0.1$ & $600$ & $120$ \\
\hline
 b) & $203.9$ &  $114.2$ & $-0.429$ & $-0.1$ & $450$ & $-90$ \\
\hline
 c) & $159.2$ &  $108.9$ & $0.398$ & $0.2$ & $100$ & $-520$ \\
\hline
 d) & $116.9$ &  $215.5$ & $-0.621$ & $0.15$ & $130$ & $676$ \\
\hline
\end{tabular}
\caption{The parameters used in the general models of Fig.~\ref{fig:voTgen}.}
\label{tab:gen_ex}
\end{center}
\end{table}

\section{Conclusions \label{sec:conclusions}}

Our analysis of strong EWPhTs in the SM plus singlet highlights the richness of possibilities this simple extension of the SM offers. By relying on a simple mean-field approximation to the finite temperature scalar potential and a judicious choice of parametrization we have been able to perform a thorough analytical study of strong EWPhTs triggered by tree-level dynamics. We have
given a strategy, summarized in Table~\ref{Table1}, to identify the regions of parameter space that would lead to such strong EWPhTs. At the same time, our analytical approach has improved the understanding of the mechanisms behind such transitions allowing us to uncover new scenarios not appreciated before. One interesting example are those transitions that rely on the presence of a flat direction in the potential at the critical temperature $T_c$,
mechanism that could operate in many other models besides the SM plus singlet one.

We have computed the important ratio $v(T_c)/T_c$ both in particular realizations of the SM plus singlet model previously studied in the literature and in some representative examples of the most general model. In models with $\mathbf{Z}_2$ symmetry we have determined that a truly strong EWPhT based on a tree-level barrier must proceed from an EW symmetric vacuum that breaks the $\mathbf{Z}_2$ symmetry to an EW broken vacuum that is $\mathbf{Z}_2$ symmetric. In other particular models and in general, we have shown that $v(T_c)/T_c$  can be easily larger than 1. This is a necessary requirement for  successful electroweak baryogenesis (switching off in the broken phase sphaleron processes that would erase the created baryon asymmetry). This jump in $v(T)$ is necessarily associated with a jump in the singlet VEV, which can also be relevant in some baryogenesis mechanisms. In addition such strong EWPhTs could also lead to a relic stochastic background of gravitational waves.

We have refined our analysis going beyond the mean-field approximation by including one-loop effects, with appropriate renormalization conditions at $T=0$ and inclusion of thermal effects with daisy resummation and no high-temperature expansions. Although $v(T_c)/T_c$ is lowered by such refinement one still finds strong EWPhTs.

As a byproduct, our parametrization of the scalar potential of the SM plus singlet, which allows a good control over its vacuum structure, can be useful also at $T=0$. In fact, the conditions summarized in the lower part of Table~\ref{Table1} can be applied at $T=0$ to guarantee the stability of the EW vacuum and then our parametrization can be applied to phenomenological analyses allowing a direct control over physical quantities. 

\newpage
\appendix
\section{Scalar Potential at Finite Temperature\label{app:thermalpot}}

\vspace{0.5cm}
\noindent
{\bf One-Loop $\mathbf{T=0}$ Corrections}  

The analysis of thermal corrections to the scalar potential requires for consistency
the inclusion of one-loop corrections to the potential at $T=0$. These are given by
the usual Coleman-Weinberg correction \cite{CW}
\be
V_1(h,s)=\frac{1}{64\pi^2} \sum_\alpha N_\alpha M^4_\alpha(h,s)\left[\log\frac{M^2_\alpha(h,s)}{Q^2}-C_\alpha\right]\ ,
\ee
where $\alpha$ runs over all degrees of freedom (counted by $N_\alpha$, which
includes a minus sign for fermions) with squared-masses $M^2_\alpha(h,s)$ (which 
depend on the scalar background fields $h$ and $s$), $C_\alpha$ is a constant
(equal to 3/2 for scalars and fermions and to 5/6 for gauge bosons) and 
$Q$ is the renormalization scale, that can be fixed {\em e.g.} to the top mass. To this we add a counter-term potential
\be
\label{Vcount}
\delta V= -\frac{1}{2}\delta \mu_h^2 h^2+\frac{1}{4}\delta \lambda_h h^4 +\frac{1}{2}\delta \mu_s^2 s^2 +
\frac{1}{4}\delta \lambda_s s^4+\frac{1}{4}\delta \mu_m s h^2+\frac{1}{4}\delta \lambda_m s^2h^2 
+\delta \mu_1^3 s+\frac{1}{3}\delta \mu_3 s^3+\delta V_0\ ,
\ee
specifically chosen so as to maintain the main properties of the tree-level potentials 
at $T=0$ derived in the text. Those potentials have two minima, the one that breaks
the electroweak symmetry at $(v_{EW},w_{EW})$ and a symmetric minimum at
$(0,w_{0,EW})$. The renormalization conditions we use cannot be used for potentials without this structure. In order to avoid problems with infrared divergent Goldstone
contributions, we find convenient to remove the Goldstone corrections
to $V_1$ in the renormalization conditions that follow (and we indicate
this by writing ${\tilde V}_1$). This is simply a change of renormalization conditions and the shift it causes in the potential shape is negligible. For alternative treatments of this complication, see {\em e.g.} \cite{Christophe}.

Explicitly, the renormalization conditions that we use are:
\be
\left.\frac{\partial (\tilde{V}_1+\delta V)}{\partial h}\right|_b=0\ , \quad
\left.\frac{\partial (\tilde{V}_1+\delta V)}{\partial s}\right|_b=0\ ,\quad
\left.\frac{\partial (\tilde{V}_1+\delta V)}{\partial s}\right|_s=0\ ,
\ee
(with the subindex $b$ indicating evaluation at the broken minimum and the subindex $s$ evaluation at the symmetric minimum)
to ensure that the one-loop minima are still located at their tree-level positions;
\be
\left.\frac{\partial^2 (\tilde{V}_1+\delta V)}{\partial h^2}\right|_b=0\ , \quad
\left.\frac{\partial^2 (\tilde{V}_1+\delta V)}{\partial h\partial s}\right|_b=0\ ,\quad
\left.\frac{\partial^2 (\tilde{V}_1+\delta V)}{\partial s^2}\right|_b=0\ ,
\ee
so that the tree-level mass matrix in the broken minimum is not affected at one-loop; then
\be
\left.\frac{\partial^3 (\tilde{V}_1+\delta V)}{\partial s^3}\right|_b=0\ ,
\ee
to have the same singlet cubic coupling;
and finally
\be
(\tilde{V}_1+\delta V)|_b=0\ ,\quad
(\tilde{V}_1+\delta V)|_s=0\ ,
\ee 
so that the one-loop values of the potential at the minima are the same as the tree-level ones. These 9 conditions allow us to determine the 9 counterterms in $\delta V$ as (dropping everywhere the subindex $EW$ for simplicity)
\bea
\delta\mu_h^2 & = & \frac{1}{2 v^2} \left\{\left[
(3-5x_w^2)\deh-(1-x_w^2)\dhh
+x_w(3x_w-2)\dhs\right]\Vob\right.\nonumber\\
&&-\left.2 x_w^2\left(3\ds-\dss+\frac{1}{6}\dsss\right)\Vob
-2 x_w^2\left(\dsVs-4\dVbs\right)
\right\}\ ,
\\
\delta\mu_s^2  &=&-\frac{1}{2\Delta w^2} \left\{
(1-3x_w^2)\left[(5\deh-\dhh)\Vob-8\dVbs\right]
+2(1-6x_w^2)\dsVs\right.\nonumber\\
&&\left.+\left[3(1-2x_w^2)(2\ds-\dhs)+\frac{1}{3}(1-6x_w +6x_w^2)\dsss\right]\Vob\right\}\ ,
\\
\delta\mu_m & = &-\frac{2}{v^2\Delta w } \left\{\left[x_w
\left(5\deh-\dhh+9\ds-2\dss+\frac{1}{3}\dsss\right)+
(1-3x_w)\dhs\right]\Vob\right.\nonumber\\
&&\left.+  x_w\left(2\dsVs-8\dVbs\right)\right\}\ ,
\\
\delta\mu_3 &=&-\frac{3}{2\Delta w^3 } \left\{\left[x_w
(5\deh-\dhh+4\ds)+\frac{1}{3}(1 -2x_w)\dsss\right]\Vob\right.
\nonumber\\
&-&\left.2x_w
\left(\dsVs+4\dVbs\right)\right\}\ ,\\
\delta\mu_1^3  &=& \frac{3}{2\Delta w } \left\{x_w\left[\frac{1}{4}(2-x_w)\dsVs
-8(1 -x_w)\dVbs\right]+
x_w(1 -x_w)(5\deh-2\dhh)\Vob\right.\nonumber\\
&&+\left.\left[\frac{x_w}{3}(1-x_w+2x_w^2)\dsss+(1-x_w)(1-2x_w)(\dhs-2\ds)\right]\Vob
\right\}\ ,
\\
\delta\lambda_h & =&\frac{1}{v^3}(\deh-v\dhh)\Vob\ , 
\\
\delta\lambda_m &=&
\frac{1}{v^2\Delta w^2}\left[\left(5 \deh- \dhh  - 3 \dhs  + 
   6 \ds - 2 \dss +\frac{1}{3} \dsss\right)\Vob+ 2 \dsVs - 8\dVbs\right]\ ,\nonumber\\
&&\\
\delta\lambda_s &=&\frac{1}{2\Delta w^4}\left[\left(5 \deh- \dhh  - 2 \dhs 
+  6 \ds  -\frac{2}{3} \dsss\right)\Vob + 4 \dsVs - 8\dVbs\right]\ ,
\\
\delta V_0 &=&\frac{1}{3}\left\{
-2x_w^2[(1-x_w^2)\dsVs-2(2-x_w^2)\dVbs]-4 {\tilde V}_{1b}
\right.\\
&&\left.+(1-x_w)^2\left[
\frac{1}{2}(1+x_w)^2(5\deh-\dhh)+x_w (2+x_w)(2\ds-\dhs)-\frac{1}{3}x_w^2\dsss\right]\Vob
\right\}\ ,\nonumber
\eea
where $\Delta w\equiv w-w_0$, $x_w\equiv w/\Delta w$, $\dVbs\equiv  {\tilde V}_{1b}-{\tilde V}_{1s}$
and $\partial_{h_r}\equiv v\partial/\partial h$, $\partial_{s_r}\equiv \Delta w \partial/\partial s$ .

These counterterms have a finite $w\rightarrow 0$ limit, do not suffer from singularities ($\Delta w$ cannot vanish), and do not spoil the $\mathbf{Z}_2$
symmetry when that is a symmetry of the Lagrangian. As explained in the text, 
strong transitions in this  $\mathbf{Z}_2$-symmetric scenario require $w=0$.
The form of the counterterms is much simpler in that case and can be obtained from the general formulas above simply setting to zero $x_w$ and all odd $s$-derivatives of $\tilde{V}_1$ evaluated at the broken minimum. In particular,
one gets $\delta \mu_m=\delta \mu_3=\delta \mu_1^3=0$.

In the model discussed in Section~\ref{sec:dark} the tree-level potential
does not have two degenerate minima but rather a nearly-flat direction, so that we do not use the previous prescription. In order to keep the nearly-flat structure at one-loop, we use instead the following renormalization conditions. For the broken minimum we impose that at one-loop it stays at the same tree-level location and with the same spectrum:
\be
\left.
\left\{\frac{\partial}{\partial h},\frac{\partial}{\partial s},
\frac{\partial^2}{(\partial h)^2},\frac{\partial^2}{\partial h\partial s},\frac{\partial^2}{(\partial s)^2}\right\}(\tilde{V}_1+\delta V)\right|_b=0\ .
\ee
The (tree-level) line $D_h^2(s)$ cuts $h=0$ at $s=s_h$ and, to maintain this at one-loop we
impose 
\be
\left.\frac{\partial(\tilde{V}_1+\delta V)}{\partial (h^2)}\right|_{(0,s_h)}=0\ .
\ee
Finally, the (tree-level) line $D_s^2(s)$ cuts $h=0$ at $s=0$ and, to maintain this and the same slope at one-loop we impose 
\be
\left.\frac{\partial(\tilde{V}_1+\delta V)}{\partial s}\right|_0=0\ ,\quad
4\mu_s^2\left.\frac{\partial^2(\tilde{V}_1+\delta V)}{\partial (h^2)\partial s}\right|_0=\mu_m
\left.\frac{\partial^2(\tilde{V}_1+\delta V)}{(\partial s)^2}\right|_0
\ee
where the subindex $0$ indicates evaluation at $h=0, s=0$.

\vspace{0.5cm}
\noindent
{\bf Finite Temperature Potential.}  

At the  large temperatures of the early Universe plasma, high-temperature effects 
modify the Higgs effective potential (or rather, free-energy). The contribution of the different plasma species to the potential, in the non-interacting gas
approximation, is given by standard one-loop (bosonic/fermionic) thermal
integrals. Each particle species, labelled by $\alpha$, contributes to the potential
\bea
\delta_\alpha V_T(h,s)&=&\frac{T^4}{2\pi^2}N_\alpha \int_0^\infty
dx\ x^2\log\left[1\pm
  e^{-\sqrt{x^2+M_\alpha^2(h,s)/T^2}}\right]
\nonumber\\
&+&\frac{T}{12\pi}\delta_{\alpha b}N_\alpha \left[M^3_\alpha(h,s) -
M^3_{T,\alpha}(h,s,T)  \right]\
, \label{Tpot}
\eea
where $M_{T,\alpha}(h,s,T)$ is the thermally corrected mass of the corresponding
species and the plus (minus) sign in the integrand is for fermions (bosons).  The second line in (\ref{Tpot}) is present only for bosons (we represent
this symbolically by writing $\delta_{\alpha b}$) and takes into account the effect of
resumming hard-thermal loops for Matsubara zero modes. For our numerical
work we used a series expansion of these integrals in terms of modified
Bessel functions \cite{AH}, avoiding high-$T$ expansions.

The thermal masses in the SM plus singlet model are as in the Standard Model except for the scalar sector. The squared mass matrix for $h$ and $s$ is
\bea
{\cal M}^2&=&
\left[
\begin{array}{cc}
-\mu_h^2+3\lambda_h h^2 +\frac{1}{2}(\mu_m + \lambda_m s)s
&\frac{1}{2}(\mu_m + 2\lambda_m s)h\\
\frac{1}{2}(\mu_m + 2\lambda_m s)h&
\mu_s^2+3\lambda_s s^2 +2\mu_3 s +\frac{1}{2}\lambda_m h^2\end{array}
\right]
\nonumber\\
&&+\frac{1}{48}\left[
\begin{array}{cc}
9g^2+3{g'}^2+2(6h_t^2+12\lambda_h+\lambda_m)&0\\
0&4(2\lambda_m+3\lambda_s)
\end{array}
\right]T^2\ .
\eea
The thermally corrected mass for Goldstones is 
\be
m_G^2 = -\mu_h^2 + \lambda_h h^2+\frac{1}{2}(\mu_m+\lambda_m s)s
+\frac{1}{48}\left[
9g^2+3{g'}^2+2(6h_t^2+12\lambda_h+\lambda_m\right]T^2\ .
\ee

\section*{Acknowledgments}

\noindent
J.R.E. and F.R. thank the CERN TH-Division for partial financial support and hospitality during the early stages of this work.  We acknowledge support from the Spanish 
Ministry MICINN under contracts FPA2010-17747 and FPA2008-01430; the Spanish Consolider-Ingenio 2010 Programme CPAN (CSD2007-00042); and the Generalitat de Catalunya grant 2009SGR894.

\end{document}